\documentclass[aip,jcp,reprint]{revtex4-1}
\usepackage[version=3]{mhchem} 
\usepackage{amsmath,amsthm,amssymb,amsfonts}
\usepackage{float}
\usepackage{siunitx}
\usepackage{graphicx}
\usepackage{booktabs}
\usepackage[english]{babel}
\usepackage{textgreek}
\usepackage{hyperref}
\usepackage[utf8]{inputenc}
\usepackage[T1]{fontenc}
\usepackage{mathptmx}
\usepackage{etoolbox}
\usepackage{dcolumn}
\usepackage{bm}

\DeclareSIUnit\angstrom{\text {Å}}
\newcommand{\dependent}{\not\mathbf{\!\perp\!\!\!\perp}}

\begin{document}


\title{Towards a robust approach to infer causality in molecular systems satisfying detailed balance\vspace{0.5cm}}

\author{Vittorio Del Tatto}
\thanks{These two authors contributed equally to this work.}
\affiliation{Scuola Internazionale Superiore di Studi Avanzati (SISSA), Via Bonomea 265, 34136 Trieste, Italy}

\author{Debarshi Banerjee}
\thanks{These two authors contributed equally to this work.}
\affiliation{Scuola Internazionale Superiore di Studi Avanzati (SISSA), Via Bonomea 265, 34136 Trieste, Italy}
\affiliation{International Centre for Theoretical Physics (ICTP), Strada Costiera 11, 34151 Trieste, Italy}

\author{Ali Hassanali}
\email{ahassana@ictp.it}
\affiliation{International Centre for Theoretical Physics (ICTP), Strada Costiera 11, 34151 Trieste, Italy}

\author{Alessandro Laio}
\email{laio@sissa.it}
\affiliation{Scuola Internazionale Superiore di Studi Avanzati (SISSA), Via Bonomea 265, 34136 Trieste, Italy}
\affiliation{International Centre for Theoretical Physics (ICTP), Strada Costiera 11, 34151 Trieste, Italy}




\begin{abstract}
The ability to distinguish between correlation and causation of variables in molecular systems remains an interesting and open area of investigation. 
In this work, we probe causality in a molecular system using two independent computational methods that infer the causal direction through the language of information transfer.
Specifically, we demonstrate that a molecular dynamics simulation involving a single Tryptophan in liquid water displays asymmetric information transfer between specific collective variables, such as solute and solvent coordinates. 
Analyzing a discrete Markov-state and Langevin dynamics on a 2D free energy surface, we show that the same kind of asymmetries can emerge even in extremely simple systems, undergoing equilibrium and time-reversible dynamics. 
We use these model systems to rationalize the unidirectional information transfer in the molecular system in terms of asymmetries in the underlying free energy landscape and/or relaxation dynamics of the relevant coordinates. 
Finally, we propose a computational experiment that allows one to decide if an asymmetric information transfer between two variables corresponds to a genuine causal link.
\end{abstract}

\maketitle

One of the most 
intriguing foundational questions in molecular science concerns how cause-and-effect relationships, plainly observed in our mesoscopic and macroscopic world, emerge from dynamic equations that are time-reversible at the microscopic scale.
Measuring
causality in a system described by classical equations of motion is highly non-trivial, and has been the object of intense investigation \cite{gorecki2006causal, kamberaj2009extracting, hacisuleyman2017entropy, hacisuleyman2017causality, sogunmez2022information, hempel2020coupling, barr2011importance, qi2013quantification, zhang2014structure, sobieraj2022granger, zhu2022neural, dutta2017spatiotemporal, perilla2013molecular, jo2015preferred}.
For this purpose, molecular dynamics (MD) simulations offer 
the possibility to interrogate specific microscopic degrees of freedom that are often unattainable or very challenging to observe and manipulate in experimental approaches.
In order to infer causality among Collective Variables (CVs) of interest, such as inter-residue distances \cite{sobieraj2022granger} or dihedral angles \cite{dutta2017spatiotemporal,sogunmez2022information} in small proteins, various studies have previously analyzed MD simulations using Granger causality (GC) \cite{granger1969investigating, shojaie2022granger},
Transfer Entropy (TE) \cite{schreiber2000measuring,schindler2007causality} or time-lagged two-body cross-correlation functions (CCFs) \cite{dutta2017spatiotemporal,hacisuleyman2017entropy}.

In other contexts, such as medical studies \cite{wu2024causal}, sociology \cite{gangl2010causal}, and epidemiology \cite{rothman2005causation}, causal questions have been addressed for decades through the lens of causal inference \cite{pearl2009causality,spirtes2016causal,runge2023causal_inference}.
This field provides a rigorous statistical framework to answer counterfactual questions such as ``Had the value of variable $X$ been different, would the value of $Y$ have been different as well?''. 
The common strategy to address this question from observational time series, namely in the absence of ad-hoc manipulations of the putative causal variable $X$, is measuring conditional dependencies between pairs of variables at different times \cite{runge2018causal}.
Informally, if a variable $Y$ at time $\tau$ depends on a variable $X$ at time zero for all possible conditioning sets including the observed variables up to time $\tau$, one infers that $X$ causes $Y$.
Crucially, this conclusion can be drawn only if no unobserved common cause of $X$ and $Y$ exists, or if all common causes of $X$ and $Y$ are included in the search space of conditioning sets.
A common cause of two variables is typically referred to as ``confounder'', and the hypothesis that all confounders are observed is referred to as ``causal sufficiency''.
While the most general algorithms to infer the causal graph rely on iterative conditional independence tests for each pair of variables \cite{spirtes1991algorithm,verma2022probabilistic,spirtes1993causation}, an alternative approach in the case of time series is to compute the Transfer Entropy (TE), which is equivalent to carry out a single conditional independence test for each pair of variables \cite{runge2018causal}.
In the following, we will refer to any measure of conditional (in)dependence in observational time series data, such as TE, as \emph{information transfer}.

If the existence of unobserved common drivers cannot be ruled out, one may assess the existence of a causal relationship $X\rightarrow Y$
by measuring the average or distributional changes in $Y$ resulting from two or more manipulations, or interventions, over $X$.
A (hard) intervention on $X(0)$, denoted as $do(X(0)=x)$, is an ideal experiment where the value of $X$ at time zero is set to $x$ independently of the value of any other variable, observed or not, that is not caused by $X(0)$.
Given two independent interventions $do\left(X(0)=x\right)$ and $do\left(X(0)=x'\right)$, the causal effect of $X(0)$ on $Y(\tau)$ can be measured 
from the difference between the post-interventional distributions $p(Y(\tau)|do(X(0)=x))$ and $p(Y(\tau)|do(X(0)=x'))$.
Importantly, this interventional approach not only allows to formulate causal statements when unobserved common drivers are present, but also provides a direct quantification of causal effects \cite{runge2019detecting}.
Pearl's ``do-calculus'' \cite{pearl2009causality} provides tools to compute post-interventional distributions from observational data, under the assumption of causal sufficiency.

In this work we investigate the emergence of \emph{strongly asymmetric} information transfers, which indicate candidate \emph{unidirectional} causal relationships, in molecular systems where the microscopic interactions are bidirectional due to Newton's third law.
In particular:

\begin{itemize}
    \item Using an extremely simple molecular system, a Tryptophan (TRP) molecule solvated in water, we show that unidirectional information transfers between one-dimensional CVs can be inferred by measuring the Transfer Entropy \cite{gorecki2006causal,kamberaj2009extracting,hacisuleyman2017entropy,hacisuleyman2017causality,hempel2020coupling,barr2011importance,perilla2013molecular,qi2013quantification,zhang2014structure,jo2015preferred,sogunmez2022information}, or using an approach introduced by some of us \cite{deltatto2024robust} when the CVs are high-dimensional (Sec.~\ref{sec:molecular_IG}).
    

    \item We show that such unidirectional information transfer can be present even in model systems which rigorously obey a time-reversible dynamics with stationary probability measure, such as a discrete-time Markov process, for which the Transfer Entropy can be computed analytically, and a Langevin dynamics on a two-dimensional potential energy surface (Sec.~\ref{sec:modelsys}).
    
     
    \item We show that \emph{if all the variables are observed}, these asymmetries allow predicting the effect of suitable interventional experiments, for example a $do(X)$, in the language of causal inference. However, the presence of a unidirectional information transfer is not a sufficient condition to decide if a causal relationship exists: in a Langevin model with three variables \emph{in which only two are observed}, we measure an asymmetric Transfer Entropy that does not correspond to a causal relationship. We show that this can be revealed by appropriate interventional experiments (Sec.~ \ref{sec:intevention}).
    
\end{itemize}

\section{Methods} \label{sec:methods}

We search for unidirectional information transfer between pairs of variables by estimating the Transfer Entropy and, for high-dimensional collective variables, by using the Imbalance Gain, an approach developed recently by some of us \cite{deltatto2024robust}.

The Transfer Entropy, in its bivariate formulation, quantifies how the future state of a random variable $Y$ can be better predicted given the knowledge of the current states of both $Y$ and a second variable $X$, rather than using only the present state of $Y$ \cite{schreiber2000measuring,palus2001synchronization}.
Given two time series $\{X_t\}_{t=0}^T$ and $\{Y_t\}_{t=0}^T$, 
we use the following definition of Transfer Entropy in direction $X\rightarrow Y$:
\begin{equation}\label{eq:transfer_entropy}
    \text{TE}_{X\rightarrow Y}(\tau) := I(X_0; Y_{\tau} \mid Y_0)\,,
\end{equation}
where $\tau$ is a discrete and positive time lag and $I(\cdot\,;\, \cdot \mid \cdot)$ is the conditional mutual information.
Condition TE$_{X\rightarrow Y}(\tau) > 0$ is equivalent to the conditional dependence relationship $X_0\, \dependent \,Y_\tau \mid Y_0$ (read: $Y_\tau$ depends on $X_0$ given $Y_0$), which allows stating the existence of a causal link (direct or indirect) from $X_0$ to $Y_\tau$, if $X$ and $Y$ are not affected by any common driver $Z$ (see Supp. Sec. S1).
If the same measure in the opposite direction, TE$_{Y\rightarrow X}$, is equal to zero after lag $\tau$, we say that the transfer of information from $X$ to $Y$ is (effectively) unidirectional after that time lag.

In molecular systems, the CVs describing the mesoscopic state are often intrinsically high-dimensional (for example, all the internal dihedrals of a protein molecule).
Estimating Transfer Entropies between multidimensional variables requires estimating high-dimensional probability distributions, and is therefore computationally demanding. 
When necessary, we will quantify information transfer by the Imbalance Gain (IG) \cite{deltatto2024robust}, a distance-based measure that we recently proposed to alleviate the practical limitations in computing Transfer Entropies between high-dimensional variables.

The Imbalance Gain probes conditional independence by a suitable rank statistics. 
Given a distance $d_A$, we define $r^{ij}_A$ the distance rank (or neighbour order) of $j$ with respect to $i$. 
Postulating that $d_A$ is informative with respect to a second distance $d_B$ when close points according to $d_A$ are also close according to $d_B$, the Information Imbalance \cite{glielmo2022ranking} from $d_A$ to $d_B$ is defined as
\begin{equation}\label{eq:info_imbalance}
    \Delta(d_A \rightarrow d_B) := \frac{2}{N} \langle r_B \mid r_A = 1 \rangle = \frac{2}{N^2} \sum_{i,j:\, r^{ij}_A = 1} r^{ij}_B\,,
\end{equation}
and it provides a number between 0 (maximum predictivity) and 1 (minimum predictivity).
The former case occurs when all nearest neighbor pairs in $d_A$ remain nearest neighbor pairs in $d_A$, while the latter occurs when such pairs are randomly distributed in $d_B$.
As shown in ref.~\cite{deltatto2024robust}, Eq.~(\ref{eq:info_imbalance}) can be extended to include $k$ nearest neighbors.

In the same spirit of Transfer Entropy and Granger Causality, in Ref.~\cite{deltatto2024robust} we proposed to use Eq.~\ref{eq:info_imbalance} to verify whether the prediction of $d_{Y_\tau}$ can be improved by using a ``mixed'' distance space including both variables $X_0$ and $Y_0$, rather than $Y_0$ alone.
Specifically, we translated the condition TE$_{X\rightarrow Y} > 0$ into the following inequality:
\begin{equation}\label{eq:inequality_II}
    \Delta(\alpha) := \min_{\alpha} \Delta (d_{\alpha X_0,\,Y_0} \rightarrow d_{Y_\tau}) < \Delta (d_{Y_0} \rightarrow d_{Y_\tau})\,.
\end{equation}
Equivalently, Eq.~(\ref{eq:inequality_II}) can be written as $\delta \Delta_{X\rightarrow Y} > 0$, by defining the Imbalance Gain (IG) in direction $X\rightarrow Y$ as \cite{deltatto2024robust}
\begin{equation}\label{eq:imbalance_gain}
    \delta \Delta_{X\rightarrow Y} := \frac{\Delta(\alpha=0) - \min_\alpha \Delta(\alpha)}{\Delta(\alpha=0)}\,.
\end{equation}

We note that previous studies using data from equilibrium molecular dynamics simulations have also considered asymmetries in the time-lagged two-body cross correlation 
functions $\langle X_0 Y_\tau\rangle$ to infer causal links \cite{dutta2017spatiotemporal,hacisuleyman2017entropy}. 
However such correlation functions are invariant under the exchange of $X$ and $Y$: $\langle X_0 Y_\tau\rangle = \langle X_{-\tau} Y_0\rangle = \langle X_\tau Y_0\rangle$ (the first equality holds under the assumption of stationarity, while the second follows from time-reversibility). 
Therefore, asymmetries in these correlation functions, if observed, can only be due to statistical errors, violations of time-reversibility induced by the integrator, and/or by the thermostat/barostat.
\section{Unidirectional information transfer between collective variables in a molecular system} 
\label{sec:molecular_IG}

\begin{figure*}   
  \centering
  \includegraphics[width=0.85\linewidth]{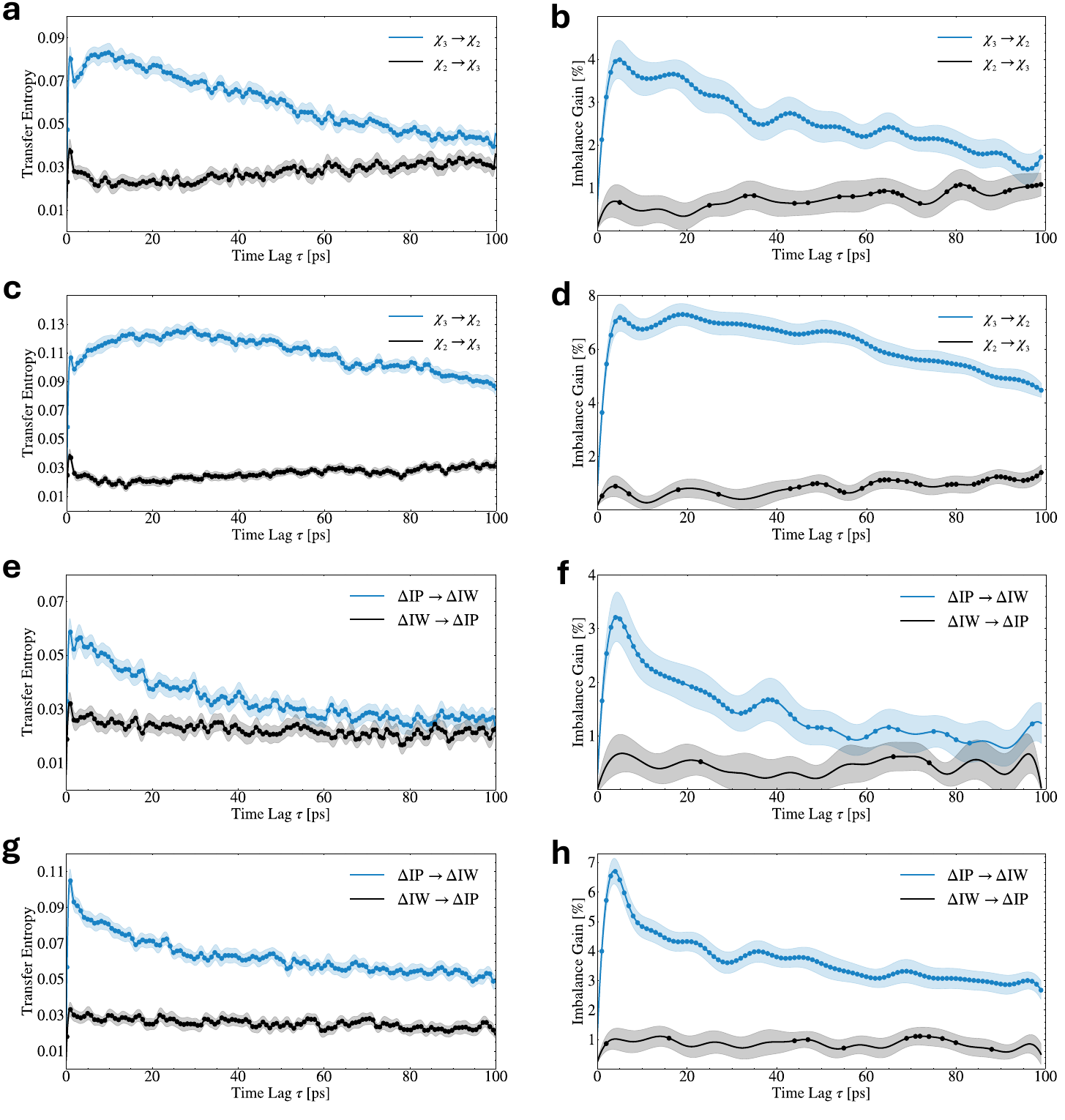}
  \caption{\textbf{a} and \textbf{b}: TE and IG curves for the CVs $\chi_{_3}$  and $\chi_{_2} $ in the GS. \textbf{c} and \textbf{d}: TE and IG curves for the same CVs in the ES. \textbf{e} and \textbf{f}: TE and IG curves for the CVs $\Delta \operatorname{IP}$ and $\Delta \operatorname{IW}$ in the GS. \textbf{g} and \textbf{h}: TE and IG curves for the same CVs in the ES. Shaded regions denote error bars over 14 independent estimates, and the bold points denote values that are determined to be significantly different from 0 using a t-test ($p < 0.001$).
  The IG was computed using $k=5$ neighbors.}
  \label{fig:IG_mol}
\end{figure*}

We first show that asymmetries in the Imbalance Gain and in the Transfer Entropy can be observed in a molecular system undergoing equilibrium and time-reversible dynamics.
These asymmetries denote unidirectional information transfer between specific collective variables and, as we will see below, candidate causal links. 
Asymmetries in the TE have already been reported in several previous studies using molecular dynamics simulations (see for example Refs. \cite{gorecki2006causal,kamberaj2009extracting,hacisuleyman2017entropy,hacisuleyman2017causality,hempel2020coupling,barr2011importance,perilla2013molecular,qi2013quantification,zhang2014structure,jo2015preferred,sogunmez2022information}). 
TE is however limited to constructing probability distributions in low-dimensions, whereas the IG was introduced precisely to overcome this limitation.

We focus on molecular dynamics simulations of the amino-acid Tryptophan (TRP) in water.
TRP is a naturally occurring fluorophore whose optical properties have been extensively studied to probe solvation dynamics - the response of protein and water coordinates following photoexcitation \cite{Nilsson_Halle_2005,Hassanali_Li_Zhong_Singer_2006,pal2002biological}. 
We conducted microsecond-long equilibrium molecular dynamics (MD) simulations of TRP in water on both the ground and excited electronic states (GS and ES, respectively) in order to uncover unidirectional dependencies between specific solute and solvent coordinates. 
Our model of excitation mirrors previous studies that involve adjusting the point charges in the indole group to capture the change in the magnitude and direction of the dipole moment \cite{Hassanali_Li_Zhong_Singer_2006, Li_Hassanali_Kao_Zhong_Singer_2007, Azizi_Gori_Morzan_Hassanali_Kurian_2023} (see Supp. Sec.~S2 for more details). 
From these simulations, we examined the relationships between a wide variety of variables that probe the coupling between the conformational changes of the TRP and the response of the surrounding water molecules. 
Figure S3 in Supp. Inf. shows a schematic of the structural coordinates that we examined. 
In addition to these structural quantities, we also examined variables that probe the changes in the interaction energy between the TRP and the environment arising from a photoexcitation. 
This can then be partitioned separately into contributions coming from the interactions of the chromophore (the indole moiety) with water molecules ($\operatorname{IW}$) and with the peptide chain ($\operatorname{IP}$).

In Fig.~\ref{fig:IG_mol} we show pairs of collective variables displaying approximately unidirectional information transfer in the TE and IG, on the timescale of the first 100 ps. 
Figures \ref{fig:IG_mol}\textbf{a} and \ref{fig:IG_mol}\textbf{b} show the TE and IG between the $\chi_{_3}$ and $\chi_{_2}$ dihedral angles in the GS, respectively, while Figures \ref{fig:IG_mol}\textbf{c} and \ref{fig:IG_mol}\textbf{d} present the same measures in the ES. 
For both sets of simulations, we observe a large IG in direction $\chi_{_3}\rightarrow \chi_{_2}$, while the IG in the opposite direction is negligible. Similar behavior is also observed in the case of the TE. In the ES, the TE and IG from $\chi_{_3}$ to $\chi_{_2}$ decay on a much longer timescale compared to the GS; in addition, we observe that the unidirectional information transfer is more marked in the ES compared to the GS. The same behavior also holds for the CVs  $\chi_{_1}$ and $\chi_{_2}$ (see Fig.~S7 in the Supp. Inf.). This suggests that the timescales associated with the flow of information between different modes is altered in the GS versus ES.

In Figures \ref{fig:IG_mol} (panels \textbf{e}, \textbf{f}, \textbf{g}, \textbf{h}), we perform the same analysis for the energetic variables $\Delta \operatorname{IP}$ and $\Delta \operatorname{IW}$.
These variables probe the total change in the electrostatic interaction energy between chromophore and peptide backbone (IP) or chromophore and water (IW), as a consequence of the excitation (see Supp. Sec.~S3). In Figures \ref{fig:IG_mol}\textbf{e} and \ref{fig:IG_mol}\textbf{f} we show the TE and IG for the energetic variables in the GS, while in Figures \ref{fig:IG_mol}\textbf{g} and \ref{fig:IG_mol}\textbf{h} we report the same measures in the ES.
In this case, we observe a unidirectional information transfer from $\Delta \operatorname{IP}$ to $ \Delta \operatorname{IW}$.
Similarly to the case of dihedrals, for the energetic variables the relaxation of both the TE and IG is significantly slowed down in the ES.  
It should be noted that for one-dimensional CVs the TE and the IG provide fully consistent results.

\begin{figure}[ht!]
  \centering
  \includegraphics[width=0.90\linewidth]{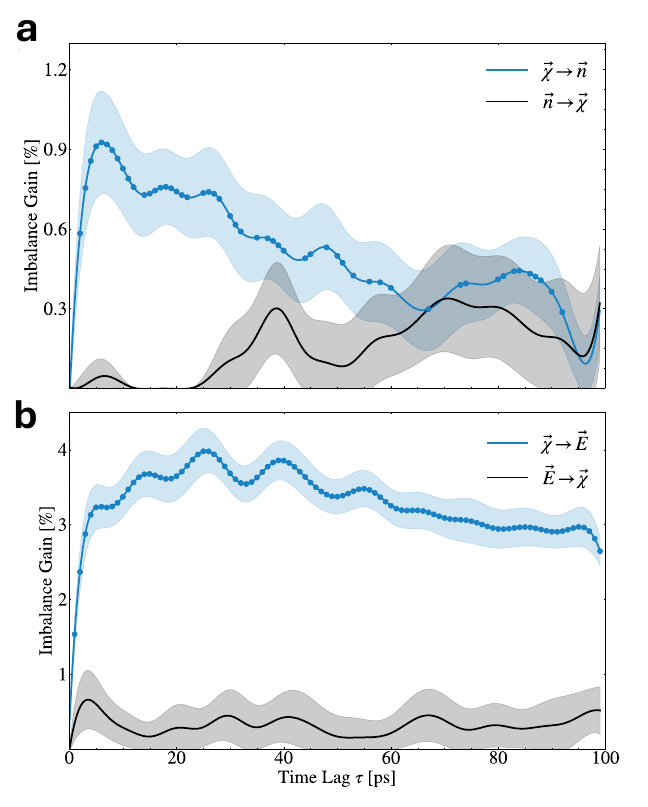}
  \caption{\textbf{a}: IG for $\vec{\chi} \rightarrow \vec{n}$ (blue) and $\vec{n} \rightarrow \vec{\chi}$ (black). \textbf{b}: IG for $\vec{\chi} \rightarrow \vec{E}$ (blue) and $\vec{E} \rightarrow \vec{\chi}$ (black). Shaded regions denote error bars over 14 independent estimates, and bold points denote values that are significantly different from 0 according to a t-test ($p<0.001$).}
  \label{fig:IG_vec}
\end{figure}

Next, we extend the analysis to multidimensional CVs, involving solvent coordinates and the energetic variables mentioned before. Conducting this type of analysis using TE is difficult due to the need to construct high-dimensional probability distributions.
Fig.~\ref{fig:IG_vec}\textbf{a} shows the Imbalance Gain between two multidimensional CVs, $\vec{\chi}$ and $\vec{n}$, which represent a collection of dihedrals and coordination numbers, respectively. The dihedral vector, $\vec{\chi}=\left(\chi_{_1},\chi_{_2},\chi_{_3}\right)$, is composed of the 3 dihedrals discussed above and shown in Fig.~S3, while $\vec{n}= \left(n_{\operatorname{CT}}, n_{\operatorname{O1}}, n_{\operatorname{O2}}, n_{\operatorname{NT}}, n_{\operatorname{NE1}} \right)$ includes the coordination numbers of the water oxygens with  the C-terminus (CT), the carbonyl O atoms (O1 and O2), the N-terminus (NT), and the indole N-H (NE1). 
Fig.~\ref{fig:IG_vec}\textbf{a} shows the emergence of a unidirectional information transfer from $\vec{\chi}$ to $\vec{n}$.
Finally, we computed the IG between $\vec{\chi}$ and $\vec{E}$, where $\vec{E} = \left(\Delta \operatorname{IP}, \Delta \operatorname{IW}\right)$, which also unveils a clear asymmetry (Fig.~\ref{fig:IG_vec}\textbf{b}).



\section{Emergence of causal links in  model systems at equilibrium}
\label{sec:modelsys}

\begin{figure*}   
  \centering
  \includegraphics[width=1.0\linewidth]{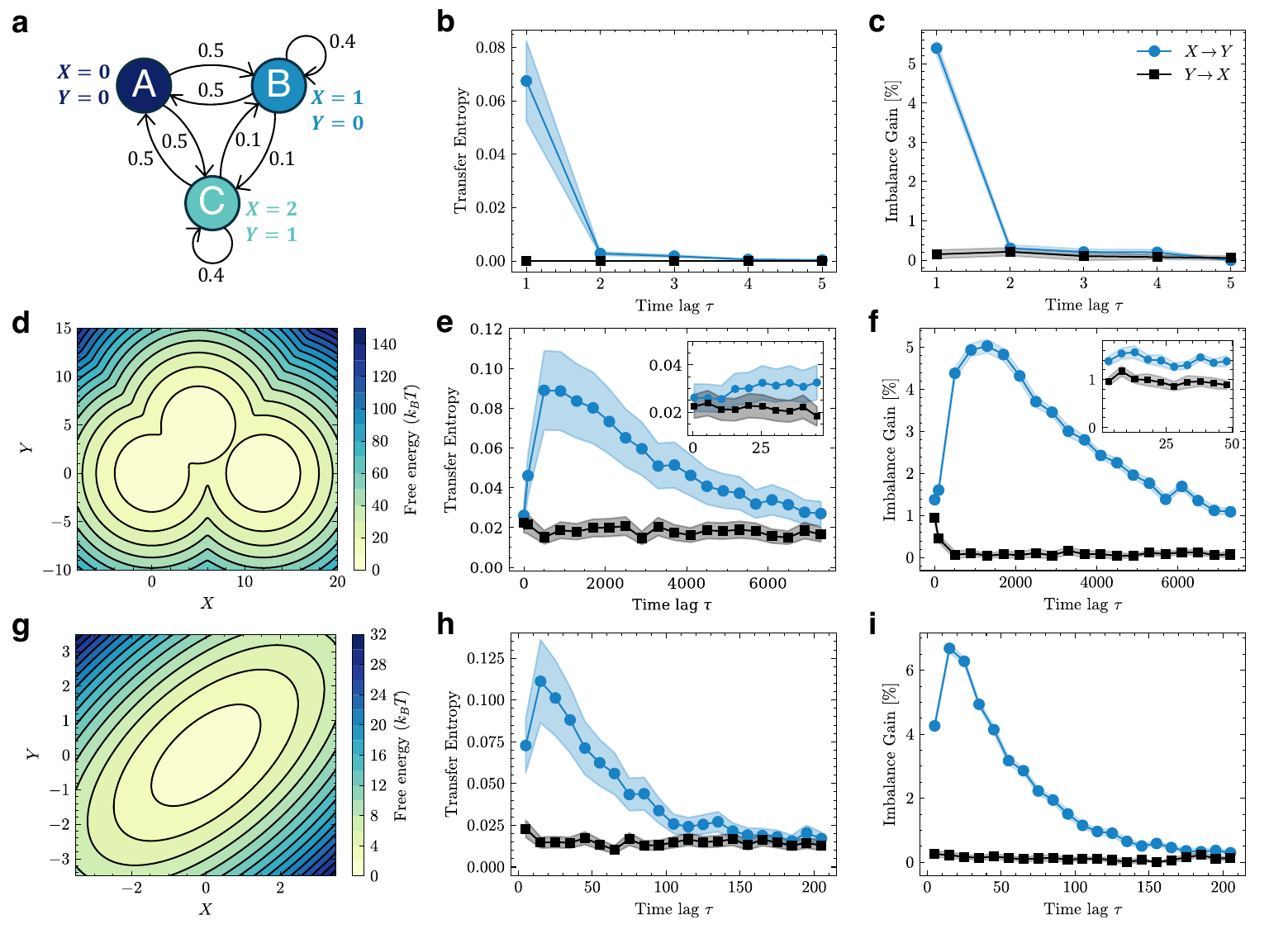}
  \caption{\textbf{a}: Markov diagram for a three-state system, with transition probabilities reported next to the arrows. $X$ and $Y$ are discrete dynamical variables whose values in each state are specified close to corresponding node.
  \textbf{b} and \textbf{c}: TE and IG from the dynamics generated by \textbf{a}, as a function of the time lag $\tau$.
  \textbf{d}: Asymmetric FES used for an overdamped Langevin dynamics with identical friction coefficients for $X$ and $Y$.
  \textbf{e} and \textbf{f}: TE and IG from the Langevin dynamics run over \textbf{d}.
  \textbf{g}: Symmetric FES used for an overdamped Langevin dynamics with different friction coefficients for $X$ and $Y$ ($\gamma_X = 1$, $\gamma_Y = 0.1$).
  \textbf{h} and \textbf{i}: TE and IG from the Langevin dynamics run over \textbf{g}. 
  Shaded regions denote error bars over 20 independent estimates, each based on $N=2000$ trajectories.
  The IG was computed using $k=25$ neighbors.}
  \label{fig:toy_systems}
\end{figure*}

To interpret the results of the previous section we conducted a similar analysis on simple model systems: a discrete-time Markov process (Fig.~\ref{fig:toy_systems}\textbf{a}) and two Langevin dynamics on different free energy surfaces (FES) (Fig.~\ref{fig:toy_systems}\textbf{d} and \textbf{g}). 
In all these systems the dynamics satisfies detailed balance. 

In the Markov system of Fig.~\ref{fig:toy_systems}\textbf{a}, the three states A, B and C are uniquely identified by variable $X$, which assumes different values (0, 1 or 2) in each state.
The second variable, $Y$, can only distinguish states A and B ($Y=0$) from state C ($Y=1$), but not A and B from each other.
Therefore, $Y$ contains information that is redundant once the value of $X$ is known.
For such a system, it is possible to show both analytically (Supp.  Sec.~S4) and numerically (Fig.~\ref{fig:toy_systems}\textbf{b}) that the TE is non-zero in direction $X\rightarrow Y$, while it is exactly zero in the reverse direction.
This finding is reproduced by the IG as a function of the time lag $\tau$ (Fig.~\ref{fig:toy_systems}\textbf{c}), which 
is significantly different from zero only in direction $X\rightarrow Y$, and only for the time lag $\tau = 1$.
We note that in this system, the actual causal link from $X$ to $Y$ is instantaneous, as $Y_t$ is a deterministic function of $X_t$.
Although TE and IG cannot directly test whether causal links are instantaneous, they can detect their presence at larger time lags (see Supp. Sec.~S1).

As a second example, we consider an overdamped Langevin dynamics (see Supp. Sec.~S5) carried out over the FES of Fig.~\ref{fig:toy_systems}\textbf{d}, which can be seen as a continuous version of the previous Markov system.
Again, variable $X$ carries more information than $Y$ about the true state of the system, as the three minima can be distinguished by projecting the free energy along $X$, while only two minima can be identified by projecting along $Y$.
This is sufficient to observe a TE unbalanced in direction $X\rightarrow Y$ (Fig.~\ref{fig:toy_systems}\textbf{e} and \textbf{f}), when time lags of order of the transition times are considered.
The information transfer in direction $Y\rightarrow X$ is instead close to zero according to both measures.
For smaller time lags, the thermal fluctuations within a single minimum play a role, and $Y$ still carries information about the state of the system that is not included in $X$. 
This is reflected by a non-zero information transfer from $Y$ to $X$ for very small $\tau$ (insets of Fig.~\ref{fig:toy_systems}\textbf{e} and \textbf{f}).

As a third example, we consider a Langevin dynamics in the FES of Fig.~\ref{fig:toy_systems}\textbf{g}, which is symmetric under the exchange of $X$ and $Y$.
If the Langevin dynamics is generated using the same friction coefficient for both the variables, no information transfer asymmetry appears between $X$ and $Y$ (see Supp. Fig.~S6).
In contrast, using a smaller friction coefficient for $Y$ 
leads to the emergence of a clear unidirectional flow from $X$ to $Y$ (Figs.~\ref{fig:toy_systems}\textbf{h} and \textbf{i}). 

Despite the differences, all previous examples describe a scenario where variable $X$ is already maximally predictive with respect to its own future, and variable $Y$ can only add redundant information on the future of $X$.
In contrast, the uncertainty over the future of $Y$ can be reduced if the current state of $X$ is known.

Importantly, in the three examples such a ``predictivity asymmetry'' emerges from different mechanisms. In the first example, $X$ provides a complete description of the system's state, while $Y$ describes the system with a certain level of degeneracy.
While $X$ resolves the degeneracy of $Y$ by distinguishing states that are identical according to $Y$, the opposite is not true.
In the second example, the system is two-dimensional, but the relaxation time within each minimum is much shorter than the transition times between the minima, so that the only information still retrievable at long time scales is the knowledge of the minimum in which the system is trapped.
Therefore, $Y$ carries non-redundant information that allows improving the prediction of $X$ only for small time lags, but not in the far future, where all relevant information is already contained in $X$.
In this case, $X$ and $Y$ are CVs retaining independent information of the true state, with $X$ being more informative than $Y$ in the long time-scale regime. 
In these first two examples, the asymmetric information transfer is rooted in the different information content that the variables retain about the true system's state.

In the third example, the symmetry of the FES implies that $X$ and $Y$ retain the same level of information of the system's state at a given time.
However, such information levels become significantly different if referred to the \emph{future} state of the system, as a consequence of the different relaxation times of $X$ and $Y$: while the description provided by $X$ at time zero can still be used as a good proxy of the system's state at time $\tau$, the same does not apply for $Y$ if $Y$ has already equilibrated.
In this scenario, the information transfer asymmetry is due to adiabatic separation: variable $X$ moves slowly, leaving to $Y$ the time to relax. 
In this condition, all the information on the long time-scale dynamics is provided by $X$ alone.



\begin{figure}[ht!]
  \centering
  \includegraphics[width=1.0\linewidth]{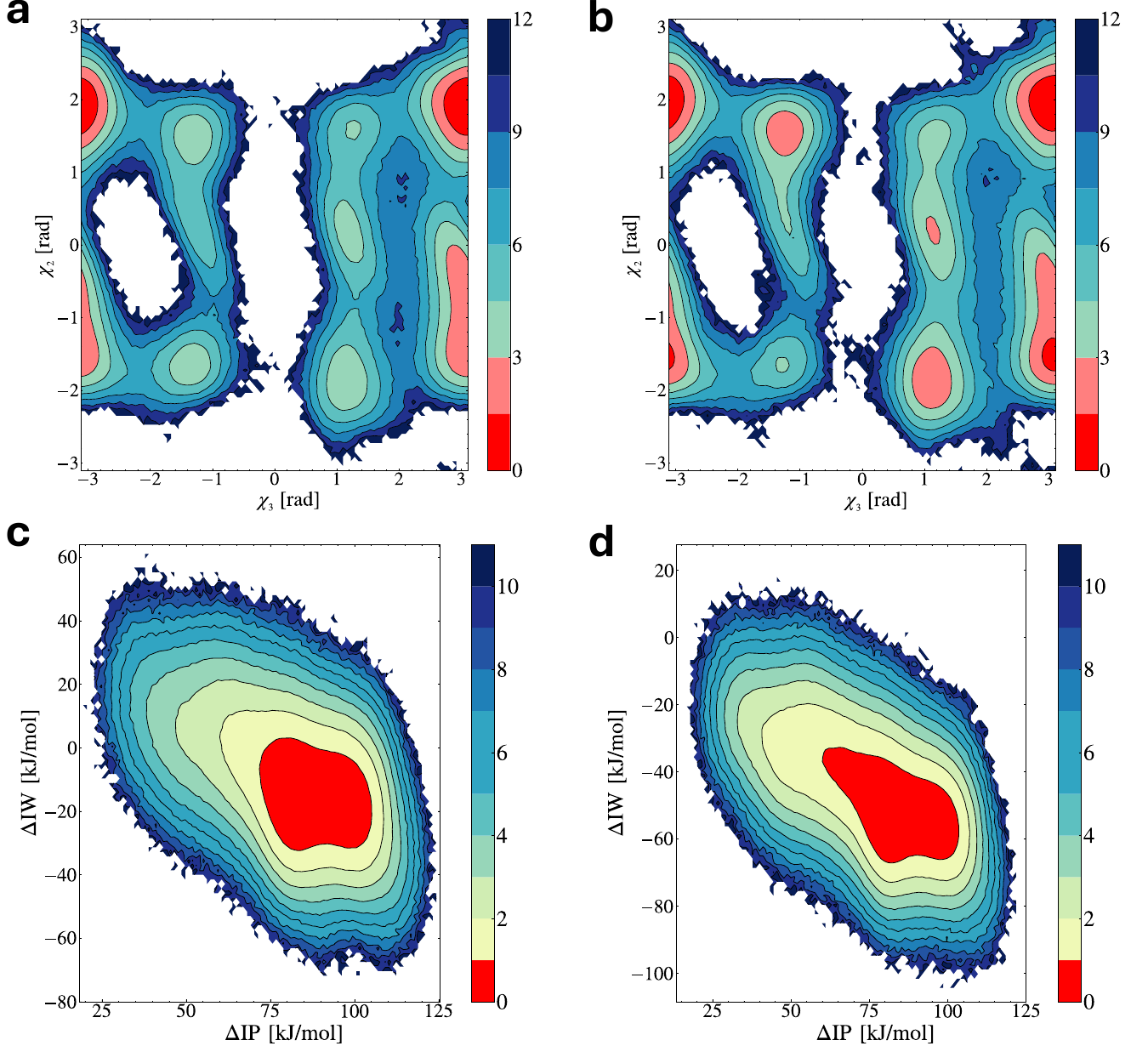}
  \caption{\textbf{a} and \textbf{b}: the FES for $\chi_{_3}$ and $\chi_{_2}$ in the GS and ES respectively. \textbf{c} and \textbf{d}: the FES for $\Delta \operatorname{IP}$ and $\Delta \operatorname{IW}$ in the GS and ES respectively.}
  \label{fig:FES_mol}
\end{figure}

The results described in Sec.~\ref{sec:molecular_IG} can be explained in light of the mechanisms just identified. In Figs.~\ref{fig:FES_mol}\textbf{a} and \textbf{b} we plot  the FES as a function of  the two dihedral angles ($\chi_3$ and $\chi_2$), for the GS and ES respectively.  
In  the FES in Fig.~\ref{fig:FES_mol}\textbf{a},  more minima can be discerned by  $\chi_{_3}$ than by $\chi_{_2}$. 
More precisely, the marginal free energies of the two angles (see Fig.~S8 in Supp. Inf.) show three minima for $\chi_{_3}$ and two minima for $\chi_{_2}$, with a higher barrier for $\chi_{_3}$ ($\sim 6-9\, k_B T$) than for $\chi_{_2}$ ($\sim 4\, k_B T$).
This indicates that $\chi_{_3}$ resolves the ``degeneracies'' of $\chi_{_2}$ more effectively than vice versa, or equivalently, that $\chi_{_3}$ serves as a better CV than $\chi_{_2}$. 
In the ES (Fig.~\ref{fig:FES_mol}\textbf{b}), some of the minima along $\chi_{_3}$ (specifically those in $(\chi_{_2},\chi_{_3})  = (-2, 1)$, $(\chi_{_2},\chi_{_3})  = (1.5, -1.25)$ and $(\chi_{_2},\chi_{_3})  = (-1.75, 3)$) become more pronounced. 
As shown in Figures ~\ref{fig:IG_mol}\textbf{c} and \textbf{d}, this leads to a more pronounced information transfer in direction $\chi_{_3}\rightarrow \chi_{_2}$. 
The slower decay of the TE and IG curves is determined by the deeper FES minima, which make $\chi_{_3}$ a slower mode in the ES than in the GS. 

To further rationalize the asymmetries observed in Fig.~\ref{fig:IG_mol}, we turned to the FES between the two energetic variables, illustrated in Figures \ref{fig:FES_mol}\textbf{c} and \textbf{d}. 
In sharp contrast to the FES involving the dihedrals, these distributions show a single broad minimum, which is only slightly asymmetric in the two variables.
However, the two CVs display decorrelation times that are different for $\Delta \operatorname{IP}$ ($\sim$ 7 ps in the GS, $\sim$ 11 ps in the ES) and $\Delta \operatorname{IW}$ ($\sim$ 1 ps in the GS, $\sim$ 2 ps in the ES), making $\Delta \operatorname{IP}$ a slower CV than $\Delta \operatorname{IW}$ (see Supp. Fig.~S9).
This leads us to conclude that the asymmetric information transfer between the indole-peptide energetics and the indole-solvent energetics is a  molecular example of the scenario shown earlier in Fig.~\ref{fig:toy_systems} (panels \textbf{g},\textbf{h},\textbf{e}).

\section{Asymmetries and the response to external interventions}
\label{sec:intevention}

\begin{figure*}[ht!]
  \centering
  \includegraphics[width=0.8\linewidth]{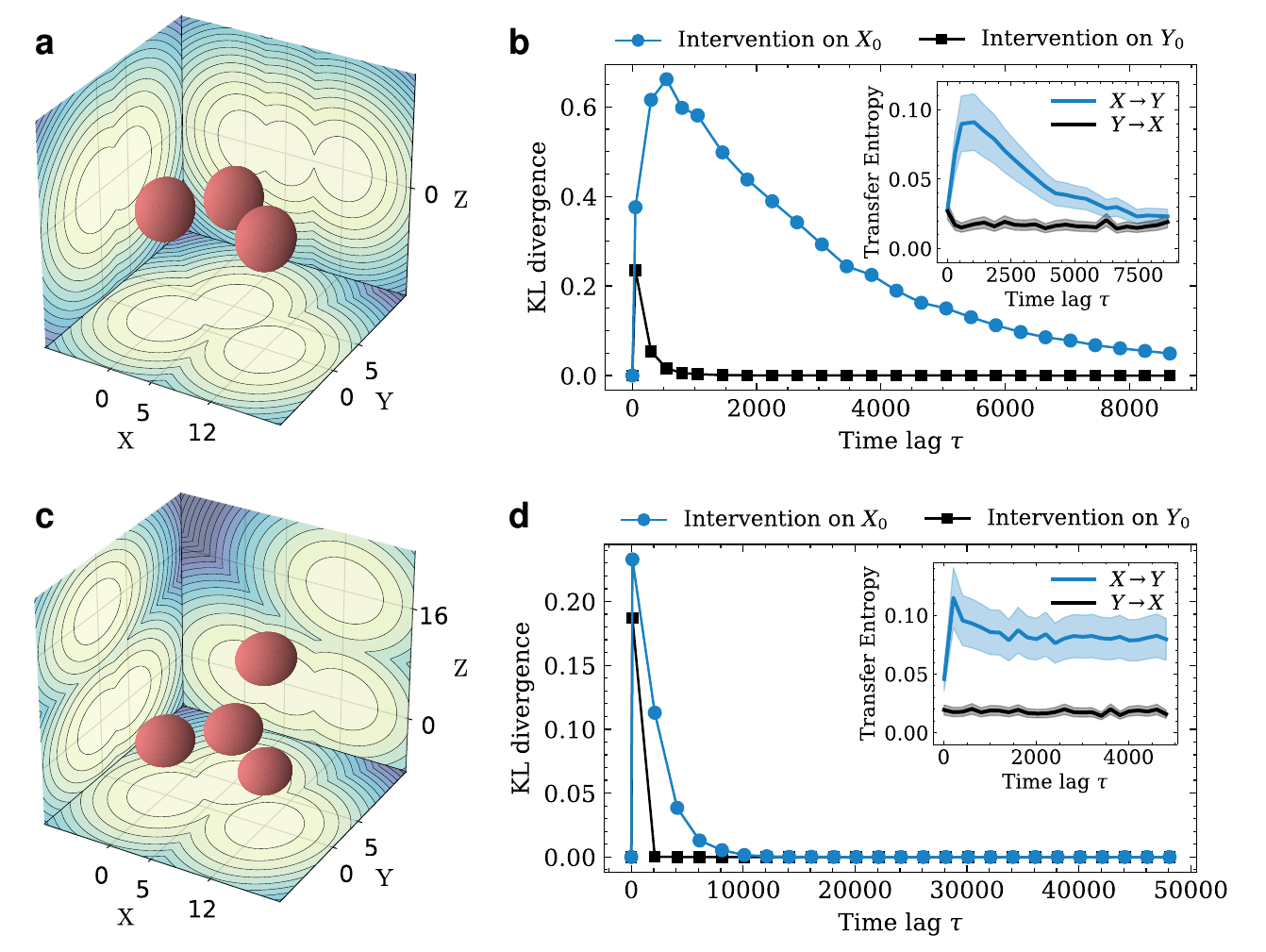}
  \caption{Effect of different interventions on two three-dimensional Langevin system with the same free energy projection in the $XY$-plane.
  \textbf{a} and \textbf{c}: Red spheres represent the free energy isosurface for $F=13\, k_B\,T$.
  The three planes show the free energy projections for all pairs variables.
  \textbf{b} and \textbf{d}: KL divergences between post-interventional distributions, intervening on $X_0$ (blue curves) and on $Y_0$ (black curves), as a function of the lag $\tau$ between the intervention and the time at which its effect is observed.
  Specifically, the two curves display $D_{KL}\left[p(Y_\tau \mid X_0 = 0) \| p(Y_\tau \mid X_0 = 12)\right]$ and $D_{KL}\left[p(X_\tau \mid Y_0 = 0) \| p(X_\tau \mid Y_0 = 5)\right]$, respectively.
  }
  \label{fig:interventions}
\end{figure*}

In this section we will show that observing a unidirectional information transfer  between two variables
is a \emph{necessary, but not sufficient} condition for the existence of a genuine causal link.
Using the tools of causal inference, we analyze the behavior of the system in response to an active intervention, which we apply by setting a variable to a specific value, without changing any other variable that is not a direct cause of the manipulated one.
In Langevin models, a ``hard'' intervention $do(X_0=x)$ can be thought to as an ideal experiment where the ``natural'' state of the system at time $t=0$, $(X_0, Y_0)$, is instantaneously set to $(x,Y_0)$.
After the external manipulation, the system is left free to evolve according to its unperturbed dynamics.
Interventions provide an intuitive framework to speak about causality: we can state that $X$ causes $Y$ if and only if an intervention on $X_0$ leads to a measurable effect on $Y_\tau$ for some $\tau > 0$.
This effect can be quantified, for example, by measuring the Kullback-Leibler (KL) divergence between the distributions of $Y_\tau$ under two different interventions on $X_0$.

In Fig.~\ref{fig:interventions} we show the effect of interventions on variables $X$ and $Y$ in two different three-dimensional model systems whose two-dimensional free energy as a function of $X$ and $Y$ is exactly identical to the free energy in Fig. ~\ref{fig:toy_systems}\textbf{d}.
We treat the third variable, $Z$, as if it were unobserved, computing only bivariate Transfer Entropies between $X$ and $Y$.

In the first system (top row) the $Z$ variable has the same distribution in the three minima, and the TE (inset in panel \textbf{b}) is qualitatively equivalent to the two-dimensional case shown in Fig.~\ref{fig:toy_systems}.
In panel \textbf{b} we show the effect of $do(X_0)$ and $do(Y_0)$ experiments (blue and black curves, respectively), using  as interventional values the positions of the furthest minima seen by each variable ($X_0=0,\,12$ and $Y_0=0,\,5$).
The effect on $Y_\tau$ of the interventions on $X_0$ is still visible for large time lags, whereas the interventions on $Y_0$ have no effect on $X_\tau$ after a time scale comparable to the relaxation time within the minima.
Therefore, in this example, the Transfer Entropy provides qualitatively the same information that one would infer by performing an external manipulation of the system.

In the second system (bottom row of Fig.~\ref{fig:toy_systems}), $Z$ does not distinguish the three minima seen by $X$, but reveals a fourth minimum, hidden for $X$, which features a higher free energy barrier than all ``visible'' barriers. 
The bivariate Transfer Entropy (inset of panel \textbf{d}) shows qualitatively the same asymmetry observed in the previous example, suggesting a unidirectional causal link $X\rightarrow Y$.
However, intervening on $X_0$ affects $Y_\tau$ on significantly shorter time scales than those deducible from Transfer Entropy (blue curve in panel \textbf{d}).
Thus, using Transfer Entropy to infer the existence of a causal effect of $X$ on $Y$ would lead to the wrong conclusion: over longer time-scales, the unobserved variable $Z$ behaves as a common driver of $X$ and $Y$. 
In Supp. Fig.~S10 we support this statement by showing that an intervention on $Z_0$ results in a long-term effect on both $X_\tau$ and $Y_\tau$, while it has no effect when applied to the system of Fig.~\ref{fig:interventions}\textbf{a}.

\section{Discussion} \label{sec:discussion}




In this work, we investigated the emergence of information transfer asymmetries in systems obeying equilibrium and time-reversible dynamics, and whether these asymmetries can be interpreted as causal links.
We measured information transfers by using the Transfer Entropy and the Imbalance Gain. Crucially, both these observables efficaciously probe three-body dependencies, involving the present state of both the putative driver and the driven variables, and the future state of the latter.
Standard time-lagged two-body cross-correlation functions (CCFs), which have been used to infer causal links from MD simulations \cite{dutta2017spatiotemporal,hacisuleyman2017entropy}, cannot report on the directional flow of information between variables in stationary and time-reversible systems, as they are symmetric by construction.
We illustrate this property by computing the CCFs between some of the relevant collective variables for TRP (see Supp. Figures~S11 and S12).

Consistent with previous studies \cite{gorecki2006causal, kamberaj2009extracting, hacisuleyman2017entropy, hacisuleyman2017causality, sogunmez2022information, hempel2020coupling, barr2011importance, qi2013quantification, zhang2014structure, sobieraj2022granger, zhu2022neural, dutta2017spatiotemporal, perilla2013molecular, jo2015preferred}, we observed empirically that information transfer asymmetries can emerge even in a simple but realistic molecular system, namely a solvated TRP molecule. The choice of this system is motivated by Fluorescence Stokes Shift experiments, where TRP can be electronically excited by absorbing UV photons and used to probe solvation dynamics \cite{trp_tdfss_jpcb_2000, trp_tdfss_pnas_2002,trp_tdfss_jacs_2006,trp_tdfss_jpcb_2015}.
Using model systems, we identified two mechanisms that explain the emergence of such asymmetries: (i) the asymmetry in the information content of different CVs, namely the capacity of one CV to describe states and transitions hidden to the others, and (ii) the discrepancy in their relaxation times. 
In particular, we found that the most informative CVs 
and the slowest CVs act as ``sources'' of information towards other CVs.
Remarkably, all the asymmetries observed in the TRP system can be explained according to either one mechanism or the other.
We also note that the first mechanism implies the second, as a CV that identifies more free energy minima can only relax on longer time-scales than a CV for which some minima are hidden.
These findings provide enhanced insight into earlier studies \cite{sobieraj2022granger,perilla2012towards}, indicating that molecular descriptors selected by Granger Causality \cite{sobieraj2022granger} or Transfer Entropy \cite{perilla2012towards} can accurately characterize transition states.
The asymmetry of information flow also opens up interesting perspectives on how to measure the chemical physics of coupling between protein and water degrees of freedom \cite{pnas_protein_slaving_solvent_2004,pnas_protein_slaving_solvent_2006,pnas_protein_slaving_solvent_2007}. For the case of TRP, we observe that there is unidirectional flow of information from protein coordinates such as the dihedrals to the solvent. It would be interesting to understand the extent to which this directionality changes for tryptophan embedded in different chemical environments in proteins.


Information transfer asymmetries inferred on equilibrium dynamics are typically associated to causal relationships \cite{kamberaj2009extracting,qi2013quantification,zhang2014structure,jo2015preferred}.
In this work, we have explicitly shown that such asymmetries are only a sufficient condition for inferring causal links, as unobserved CVs - namely, those not considered in the analysis - may act as common drivers of observed CVs.
Specifically, if $Z$ identifies a higher free energy barrier than those observed by $X$ and $Y$, $Z$ behaves as a common driver of $X$ and $Y$ on time scales comparable with the transition time to the hidden minimum.

Our findings suggest two possible routes for discovering causal relationships in molecular systems: either using a set of CVs that can be safely assumed to be ``causally sufficient" -- that is, unaffected by unobserved common drivers, or performing explicit interventional experiments on CVs of interest.
The first approach is viable by considering a large pool of CVs, such as all key dihedrals of a molecule \cite{sobieraj2022granger}, and estimating information transfers in a multivariate fashion (see Supp. Sec.~S1).
This approach is unavoidably affected by the curse of dimensionality as the number of CVs increases, although methodologies designed for high-dimensional settings, such as the IG and its extensions \cite{allione2025linear}, promise to alleviate this issue.

The second approach necessitates the design of interventional experiments on molecular CVs. This approach sounds natural in a simulation setting, in which one can perform arbitrary manipulations on the system, but poses some practical challenges.
In particular, the interventional experiments that we carried out on the model systems (Sec.~\ref{sec:intevention}) were applied 
to ``orthogonal'' CVs $X$ and $Y$, such that an instantaneous variation of $X$ at time $t=0$ does not affect $Y$ at the same time.
This may be the case for CVs that describe spatially separated regions of a molecular system, such as distant sites within a protein.
However, CVs of interest can also depend on a common subset of degrees of freedom that generate instantaneous dependencies. 
In this scenario, setting $X(0)$ to an arbitrary value ($do(X(0)=x)$) may be practically impossible without changing also $Y(0)$.
As an example, in our TRP system, the dihedrals $\chi_{_1}, \chi_{_2}$ and $\chi_{_3}$ depend on common atomic positions affected by rigid constraints, and in turn, not all arbitrary choices of such angles are possible.
Moreover, ``hard'' interventions such as those applied in this work appear challenging in molecular dynamics simulations, as they would require an instantaneous modification of several degrees of freedom, making it necessary to develop appropriate protocols.
The design of suitable interventional experiments on molecular CVs will be the subject of future work.

\begin{acknowledgments}
DB and AH thank the European Commission for funding on the ERC Grant HyBOP 101043272. DB and AH also acknowledge MareNostrum5 (project EHPC-EXT-2023E01-029) for computational resources.
This work was partially funded by NextGenerationEU through the Italian National Centre for HPC, Big Data, and Quantum Computing (Grant No. CN00000013 received by A.L.).
\end{acknowledgments}


\section*{Data Availability Statement}
The data that support the findings of this study are available from the corresponding author upon reasonable request.

\bibliography{refs}

\end{document}


\beginsupplement

\section{Transfer Entropy}\label{sec:transfer_entropy_SI}
\subsection*{Bivariate case}
In this Section we compare alternative definitions of the bivariate Transfer Entropy.
The standard definition provided by \citet{schreiber2000measuring} reads
\begin{equation}\label{eq:te_canonical_SI}
    \text{TE}_{X\rightarrow Y} = I\left(X_t^{-}; Y_{t+1} \mid Y_{t}^{-} \right)\,,
\end{equation}
where $X_t^{-}$ and $Y_t^{-}$ denote time windows built over the past of $X$ and $Y$:
\begin{subequations}\label{eqs:embeddings_SI}
    \begin{align}
        X_t^{-} &= (X_t, X_{t-1}, ..., X_{t-k+1})\,, \\
        Y_t^{-} &= (Y_t, Y_{t-1}, ..., Y_{t-l+1})\,.
    \end{align}
\end{subequations}
Equations~(\ref{eqs:embeddings_SI}) assume that the dynamics of $X$ and $Y$ can be approximately described as Markov processes of order $k$ and $l$, respectively.
Notice that this definition inherently assumes $\tau=1$.

\citet{palus2007directionality} provided a formulation with an explicit dependence on the time lag between the present and the future:
\begin{equation}\label{eq:te_palus_SI}
    \text{TE}_{X\rightarrow Y}(\tau) = I\left(X_t^{-}; Y_{t+\tau} \mid Y_{t}^{-} \right)\,.
\end{equation}
As we observed in related works \cite{deltatto2024robust,allione2025linear}, the choice of $\tau > 1$ can be beneficial for analyzing time series generated by continuous dynamics, since conditional dependencies may manifest more clearly at larger time scales compared to those of the true interactions.

The definition provided in Eq.~(1) of the main text, which we used throughout this work, is obtained from Eq.~(\ref{eq:te_palus_SI}) by replacing $X_t^{-}$ and $Y_{t}^{-}$ with the single frames $X_t$ and $Y_{t}$, respectively.
For the Markov system of Fig.~3 in the main text, the conditional mutual information appearing in the definition was computed using the Python \texttt{pyitlib} library.
In all the other applications, we employed the estimator proposed by \citet{mesner2021conditional} and implemented in the Python library \texttt{knncmi}, using $k=7$ neighbors.
For computing the Transfer Entropy between dihedral angles, the distances employed by such a nearest-neighbor estimator were computed by enforcing the correct periodicity.

\subsection*{Connection to Structural Causal Models}
We use the language of Structural Causal Models (SCMs) \cite{chicharro2014algorithms} to draw a connection between the conditional dependence relationships assessed by Transfer Entropy and the formulation of causal statements.
A discrete-time SCM for two time-dependent variables $X_t$ and $Y_t$ can be written as:
\begin{subequations}\label{eq:sce}
\begin{align}
    X_{t} &:= f(pa(X_t), \eta^X_t)\,, \label{eq:1st_sce}\\
    Y_{t} &:= g(pa(Y_t), \eta^Y_t)\,, \label{eq:2nd_sce}
\end{align}
\end{subequations}
where $f$ and $g$ are generic and possibly nonlinear functions which describe the details of the interactions, $pa(X_t)$ ($pa(Y_t)$) is a set containing all direct causes of $X_t$ ($Y_t$), called its ``parents'', while $\eta_t^X$ and $\eta_t^Y$ are noise terms which account for the effect of unobserved variables.
In this model, causal sufficiency is satisfied by requiring that $\eta_t^X$ and $\eta_t^Y$ are independent random variables, as any dependence would imply the presence of a common (and unobserved) cause of $X_{t}$ and $Y_{t}$.
Under fairly general assumptions \cite{runge2018causal}, the structure of Eqs.~(\ref{eq:sce}) is uniquely identified by a set of conditional independence relationships.
For example, condition $X_{t-\tau}\not\in pa(Y_{t+1})$ (namely, $X_{t-\tau}$ is \emph{not} a direct cause of $Y_{t+1}$) corresponds to the conditional independence relationship $X_{t-\tau}\,\indep \,Y_{t+1} \mid pa(Y_{t+1})$.
In order to directly assess such a condition from time series data, one has to either infer the conditioning set $pa(Y_{t+1})$ using iterative approaches \cite{chicharro2014algorithms,runge2018causal}, or use a larger set that includes $pa(Y_{t+1})$ without breaking the conditional independence above.
Assuming that the SCM does not include instantaneous causal links, 
this can be achieved by using as conditioning set the entire past of the ``target'' variable, $X_{t}^{-} \cup Y_{t}^{-}$, of which $pa(Y_{t+1})$ is necessarily a subset.
As shown in Eqs.~(\ref{eqs:embeddings_SI}), $X_t^{-}$ and $Y_t^{-}$ can be constructed in practice up to the maximum interaction lag that is believed to be present in the unknown dynamics.
Following this strategy, the tested relationship $X_{t-\tau}\,\indep \,Y_{t+1} \mid pa(Y_{t+1})$ can be replaced by 
\begin{equation}\label{eq:cond_indep_fullpast}
X_{t-\tau}\,\indep \,Y_{t+1} \mid \left(X_t^{-}   \cup Y_t^{-} \right)\setminus X_{t-\tau}\,,
\end{equation}
where $\left(X_t^{-} \cup Y_t^{-} \right)\setminus X_{t-\tau}$ denotes the past of the series without $X_{t-\tau}$.
If this condition is satisfied, then $X_{t-\tau}\not\in pa(Y_{t+1})$; otherwise, one concludes that $X_{t-\tau}$ is a direct cause of $Y_{t+1}$.

Transfer Entropy can be viewed as a method for testing conditional independencies of the form given in Eq.~(\ref{eq:cond_indep_fullpast}), by considering multiple time lags simultaneously.
Specifically, Schreiber's definition (see Eq.~(\ref{eq:te_canonical_SI})) tests the conditional independence $X_t^{-}\,\indep\,Y_{t+1}\mid~Y_t^{-}$, which is obtained from Eq.~(\ref{eq:cond_indep_fullpast}) by replacing $X_{t-\tau}$ with $X_t^{-}$. 
In this case, TE$_{X\rightarrow Y} > 0$ allows deducing that a causal link from a component of $X_t^{-}$ to $Y_{t+1}$ exists, but not the specific interaction lag $\tau$.
For example, the conditional \emph{dependence} $X_t^{-}\,\dependent\,Y_{t+1}\mid~Y_t^{-}$ may originate from direct links $X_{t-\tau}\rightarrow Y_{t+1}$, for any $X_{t-\tau}\in X_t^{-}$. 
Similarly, the conditional independence tested by the Transfer Entropy of \citet{palus2007directionality} (see Eq.~(\ref{eq:te_canonical_SI})) is $X_t^{-}\,\indep\,Y_{t+\tau}\mid Y_t^{-}$.
Also in this case, condition TE$_{X\rightarrow Y}>0$ allows inferring the existence of a lag-unspecific causal link from $X$ to $Y$.
Finally, the Transfer Entropy employed in this work (see Eq.~(1) in the main text) tests the conditional independence $X_t \,\indep\, Y_{t+\tau} \mid Y_t$.

Importantly, instantaneous links such as that from $X$ to $Y$ in the three-state Markov system (Fig.~3\textbf{a} of the main text) can still be detected by measuring Transfer Entropy, although not identified as instantaneous.
In that system, for example, the direct and instantaneous link $X_t \rightarrow Y_t$ generates a conditional dependence $X_t\,\indep\, Y_{t+1} | Y_t$ through the indirect causal link $X_t \rightarrow X_{t+1} \rightarrow Y_{t+1}$, which results in $\text{TE}_{X\rightarrow Y}(\tau=1) > 0$ (see Fig.~3\textbf{b} in the main text and Supp. Sec.~\ref{sec:markov_system_SI}).

\subsection*{Multivariate Transfer Entropy}
The multivariate generalization of the Transfer Entropy used in this work is:
\begin{equation}\label{eq:multiv_transfer_entropy}
    \text{TE}_{X\rightarrow Y \mid Z}(\tau) := I(X_t; Y_{t+\tau} \mid Y_t, Z_t)\,,
\end{equation}
where $Z_t$ is a (possibly multidimensional) variable containing putative common drivers of $X$ and $Y$.
Eq.~(\ref{eq:multiv_transfer_entropy}) tests the conditional independence relationship $X_t\,\indep\,Y_{t+\tau} \mid \{Y_t, \,Z_t\}$, which allows stating that $X_t$ is not a direct or indirect cause of $Y_{t+\tau}$.
Similarly to the bivariate case, such a conclusion can be drawn by assuming that there are no unobserved common drivers of $X$ and $Y$ (namely, variables causing both $X$ and $Y$ which are not included in $Z$).
The fact that this measure does not allow distinguishing between direct and indirect causal links, namely mediated by $Z$, directly comes from the fact that it does not infer the specific interaction lag, as in the bivariate case.
As we discussed in Ref.~\cite{allione2025linear}, the use of a single frame at time $t$ in place of a time window does not affect the interpretation of the measure up to the assumption that the maximum time lag of direct links is 1,  which is satisfied by all time series employed in this work.

\begin{figure}[H]
  \centering
  \includegraphics[width=1.0\linewidth]{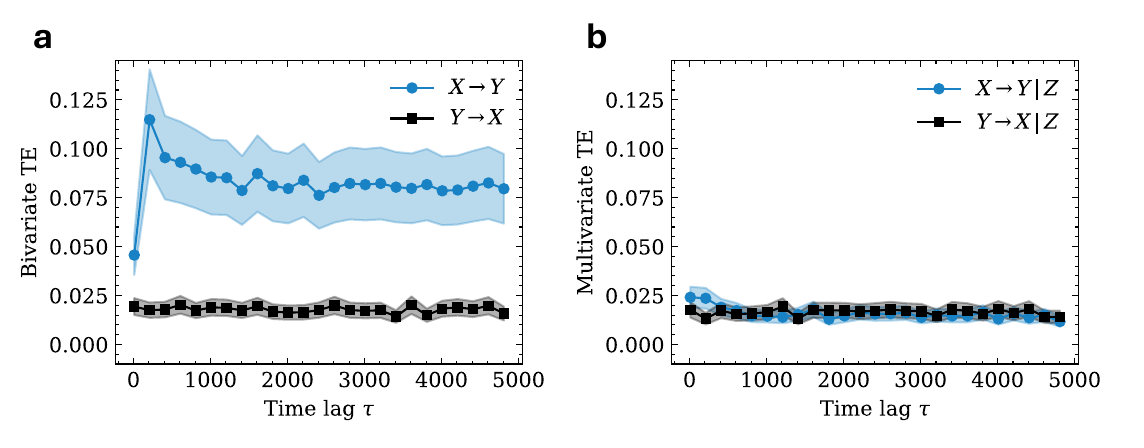}
  \caption{\textbf{a}: Bivariate TE from $X$ to $Y$ (blue) and from $Y$ to $X$ (black) as a function of the lag $\tau$, for the Langevin systems of Fig.~5\textbf{c}.
  \textbf{b}: Multivariate TE$_{X\rightarrow Y \mid Z}$ (blue) and TE$_{Y\rightarrow X \mid Z}$ (black) for the same Langevin system.}
  \label{fig:TE_biv_vs_multiv}
\end{figure}

\begin{figure}[H]
  \centering
  \includegraphics[width=1.0\linewidth]{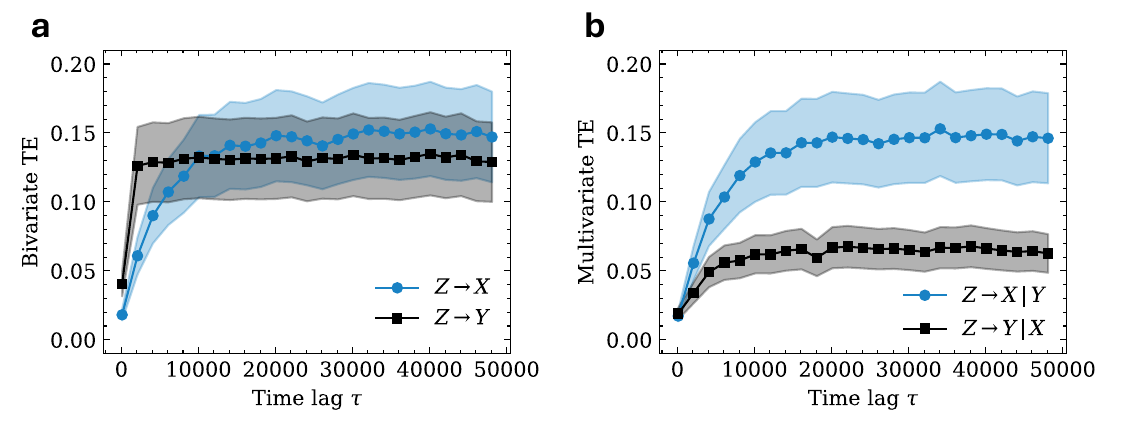}
  \caption{\textbf{a}: Bivariate TE from $Z$ to $X$ (blue) and from $Z$ to $Y$ (black) as a function of the lag $\tau$, for the Langevin systems of Fig.~5\textbf{c}.
  \textbf{b}: Multivariate TE$_{Z\rightarrow X \mid Y}$ (blue) and TE$_{Z\rightarrow Y \mid X}$ (black) for the same Langevin system.}
  \label{fig:TE_biv_vs_multiv_Z}
\end{figure}

In Fig.~\ref{fig:TE_biv_vs_multiv} we show a comparison between the bivariate TE$_{X\rightarrow Y}$ (panel \textbf{a}) and the multivariate TE$_{X\rightarrow Y \mid Z}$ (panel \textbf{a}), for the three-dimensional Langevin system shown in Fig.~5\textbf{c} of the main text.
The bivariate TE erroneously suggests the presence of a link from $X$ to $Y$ in the large time-scale regime, while its multivariate version allows deducing that $Z$ is a common driver of $X$ and $Y$.
This conclusion can be drawn by observing that TE$_{Z\rightarrow X \mid Y} > 0$ and TE$_{Z\rightarrow Y \mid X} > 0$ (Fig.~\ref{fig:TE_biv_vs_multiv_Z}).

\section{Molecular Dynamics Simulations}\label{sec:MD_simulations_SI}
We performed molecular dynamics simulation of a Tryptophan molecule in water solution.
Tryptophan (TRP) is a fluorescent amino acid which serves as a model molecular system to investigate the coupling and driving of different solvent and solute degrees of freedom. It is also an ideal starting point to model the effects of an external perturbation on a molecular system, particularly the protein-solvent interactions. In our case, this external perturbation is the photo-excitation of TRP. The indole group, which is the chromophore in TRP, has two closely spaced electronic excited states, $\operatorname{^1 L_a}$ and $\operatorname{^1 L_b}$. The lowest excited singlet state of the indole in the gas phase and non-polar solvents is $\operatorname{^1 L_b}$ state, while in polar media (such as water), the $\operatorname{^1 L_a}$ state is lowest energy singlet state \cite{Song_Kurtin_1969, Albinsson_Kubista_Norden_Thulstrup_1989, Serrano-Andrés_Roos_1996, Sobolewski_Domcke_1999, Dedonder-Lardeux_Jouvet_Perun_Sobolewski_2003}. The indole moiety can be modelled in a ``classical" fashion simulating vertical electronic excitation using a protocol based on the work of Sobolewski and Domcke \cite{Sobolewski_Domcke_1999}. The ground state topology charges generated by the force field are left unaltered, and a topology to simulate the system in the excited state is generated by applying the Mulliken charge differences between $\operatorname{^1 L_a}$ and $\operatorname{S_0}$ from the CASSCF calculations in Ref.\citenum{Sobolewski_Domcke_1999}. The ultrafast internal conversion between $\operatorname{^1 L_b}$ and $\operatorname{^1 L_a}$ which occurs in a few femtoseconds in polar solvents is not modelled in this approach. This protocol is well-established and has been used in the past for modelling the indole moiety of TRP in the excited state and studying a variety of physical interactions in good agreement with experiments \cite{Hassanali_Li_Zhong_Singer_2006, Li_Hassanali_Kao_Zhong_Singer_2007, Azizi_Gori_Morzan_Hassanali_Kurian_2023}. The veracity of this protocol can be further justified by analysing the dipole moment of the indole group, since this is a distinctive feature of the difference between the ground state and the 2 accessible excited states. Indeed, our ground state (GS) and excited state (ES) configuration has a dipole moment of 2.4 D and 4.8 D respectively, which is consistent with the values from Ref.\citenum{Sobolewski_Domcke_1999}.

The Zwitterionic form of TRP was chosen where the N-terminus was protonated as -NH3$^+$ and the C-terminus was deprotonated as -COO$^-$. The AMBER ff19SB protein force field \cite{ff19SB} was used, and as per the recommendation of the AMBER developers, it was paired with the non-polarizable four-point rigid Optimal Point Charge (OPC) water model \cite{OPC, OPC3}. Since we are interested in highly accurate protein-solvent interactions, the OPC model was an obvious choice due to the faithful reproduction of various experimental properties of bulk water, such as density, dielectric constant, heat of vaporization, self-diffusion coefficient and viscosity\cite{OPC_melting_point, OPC_surface_tension, OPC_viscosity}. GROMACS 2023.2 \cite{gmx_1995, gmx_2001, gmx_2005, gmx_2008, gmx_2013, gmx_2015_article, gmx_2015_conference, gmx_2023-2} was used to perform the simulations where the system was first equilibrated in the NVT ensemble at 300 \unit{\kelvin} for 10-\unit{\nano\second} using the Stochastic Velocity Rescaling thermostat \cite{bussi_thermostat}. Then, we run a constant pressure (NPT) simulation for a further 10-\unit{\nano\second} using the Stochastic Cell Rescaling barostat \cite{bussi_barostat} at 1 atm and 300 \unit{\kelvin}. Eventually, we carried out a production NVT run of 1-\unit{\micro\second} at 300 \unit{\kelvin} using the Leapfrog integrator with a 2-\unit{\femto\second} time-step. Buffered Verlet lists were used to keep track of atomic neighbors. The long-range electrostatic interactions were calculated using the smooth Particle Mesh Ewald (PME) algorithm \cite{pme_1993, pme_1995}. A shifted Lennard-Jones potential with a cutoff value of 1.2 \unit{\nano\metre} and long range dispersion corrections for energy and pressure was used. The system was solvated with 960 OPC water molecules in a cubic box of size 3.5 \unit{\nano\metre} with periodic boundary conditions. The geometry of the water molecules was constrained by the SETTLE algorithm \cite{SETTLE_1992} while for the TRP, bonds with hydrogen atoms were constrained using LINCS \cite{LINCS_1997}.

The final dataset for the Imbalance Gain (IG) analysis is constructed by sampling frames every 1-\unit{\nano\second} from the 1-\unit{\micro\second} trajectories in the GS and ES separately, and using these as the starting points for 1000 short 500-\unit{\pico\second} simulations in the NVE ensemble.
This ensemble is used instead of the more common NVT or NPT settings in order to emulate the conditions of the dynamical systems described in Sec.~3.
Indeed, the dynamics in the microcanonical ensemble is rigorously time-reversible, as the Hamilton's equations satisfy a well-known time-invariance property under a change of sign of the momenta.
Since a practical implementation of this dynamics is likely to violate time-reversibility for numerical and/or algorithmic reasons, as evidenced by a non-zero energy drift,
we fine-tuned the simulation parameters in order to keep this drift below $-5.95\times 10^{-5}$ kJ/(mol$\cdot$ps) per atom.
All rotations and translations were removed from the TRP from these NVE simulations, and the resultant trajectories were used for the IG and TE analysis.

\section{Definition of the collective variables}\label{sec:CVs_SI}

\begin{figure}[H]
  \centering
  \includegraphics[width=0.6\linewidth]{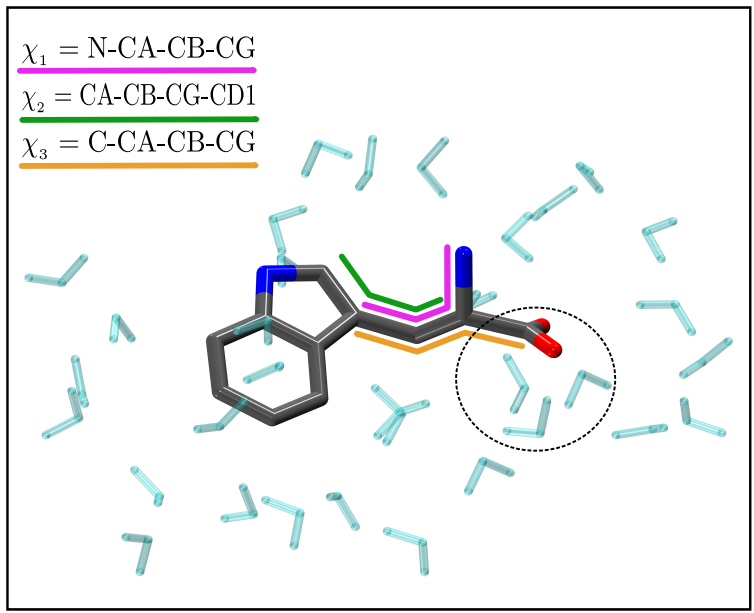}
  \caption{A TRP molecule in ``licorice'' representation with some surrounding water molecules. The three key dihedrals, $\chi_{_1}$, $\chi_{_2}$ and $\chi_{_3}$, are shown in bold lines colored in pink, green, and orange respectively. The dotted circle depicts a key region for measuring coordination numbers and protein-solvent interactions.}
  \label{fig:TRP}
\end{figure}

\subsection*{Dihedral Angles}
The three dihedral angles of interest to us are shown in Fig.~\ref{fig:TRP}: $\chi_{_1}$ shows the linkage of the N-terminus to the indole group, $\chi_{_2}$ shows the orientation of the indole with respect to $C_\alpha$, and $\chi_{_3}$ shows the linkage of the C-terminus to the indole moiety. 

\subsection*{Coordination numbers}
The relevant coordination numbers that were considered are those of OW (oxygen in water molecules) to the most polar atoms in the system: the N-terminus ($n_{\operatorname{NT}}$), the C-terminus ($n_{\operatorname{CT}}$), the 2 carbonyl O's of the C-terminus ($n_{\operatorname{O1}}$, $n_{\operatorname{O2}}$), and the N-H group of the indole ($n_{\operatorname{NE1}}$). 
These coordination numbers were calculated using a smooth switching function of the form:

\[ \sum_{i} \sum_{j} s_{ij} = \dfrac{1.0}{1.0 + \exp((d_{ij}-d_0)/\sigma)}\,, \]

where $\sigma$ controls the width of the switching function, and $d_0$ is a cutoff distance for 2 atoms $i$ and $j$. In this work, $d_0 = 0.15$ was chosen.

\subsection*{Interaction Energy gaps}

The indole-peptide interaction energy ($\operatorname{IP}$) is defined as the sum of the total Coulomb interaction (including both short-range and long-range components) between all the atoms on the indole moiety and the peptide, which is the rest of the TRP molecule. Similarly, the indole-water interaction energy ($\operatorname{IW}$) is defined as the cumulative Coulomb interaction between all the atoms on the indole moiety and the water molecules.

In our system, we simulated a vertical excitation of the system by replacing the ground-state charges with that of the excited state. This gives us the ``interaction energy gaps'' collective variables as:

\begin{equation*}
    \Delta \operatorname{IP} = \operatorname{IP}_{S_1} - \operatorname{IP}_{S_0}  
\end{equation*}
\begin{equation*}
    \Delta \operatorname{IW} = \operatorname{IW}_{S_1} - \operatorname{IW}_{S_0}  
\end{equation*}

The quantities $\Delta \operatorname{IP}$ and $\Delta \operatorname{IW}$ give us the contribution of the indole-peptide and indole-water interaction energies, respectively, to the total energy gap between the ground and excited state. This methodology allows us to probe the peptide and solvent contributions to the TDFSS. Extensive prior research has explored the use of energy gaps between ground and excited states as a probe for a wide range of chemical systems \cite{marcusTHEORYOXIDATIONREDUCTION1963,marcusTheoryOxidationreductionReactions1957,marcusTheoryOxidationreductionReactions1957a,energy_gap_Warshel_JPC_1982,energy_gap_Tachiya_JPC_1993,energy_gap_Hwang_Warshel_JACS_1987,energy_gap_King_Warshel_JCP_1990,energy_gap_Rose_Benjamin_1994,energy_gap_Small_Matyushov_Voth_JACS_2003,energy_gap_Blumberger_Tavernelli_JACS_2004,energy_gap_Blumberger_Tavernelli_JCP_2005,energy_gap_Blumberger_Tavernelli_JCP_2006,Li_Hassanali_Kao_Zhong_Singer_2007}.


\section{Asymmetric transfer entropy in the Markov system}\label{sec:markov_system_SI}

We show in this Section that TE$_{X\rightarrow Y}(\tau)>0$ and TE$_{Y\rightarrow X}(\tau)=0$ for the Markov system depicted in Fig.~3 of the main text, focusing on the case $\tau = 1$.

\begin{figure}[H]
  \centering
  \includegraphics[width=0.75\linewidth]{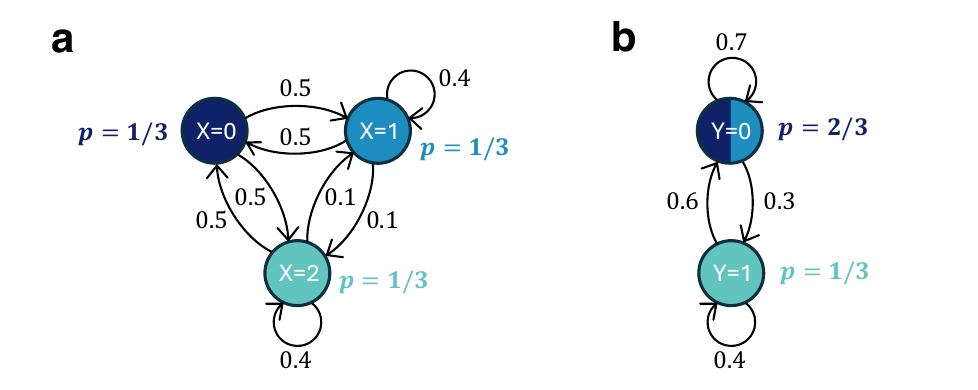}
  \caption{Marginal Markov diagrams for the variables $X$ (panel \textbf{a}) and $Y$ (panel \textbf{b}), derived from the Markov diagram of Fig.~3\textbf{a} in the main text.
  Transition probabilities are reported in black next to the arrows; probabilities of the equilibrium distribution are written next to the nodes with the same color of the corresponding state.
  While the diagram for $X$ is equivalent to the original Markov system, in the diagram for $Y$ two states are combined into a single node.}
  \label{fig:markov_SI}
\end{figure}

Denoting with $S_t$ the state of the system at time $t$ ($S_t\in\{A,B,C\}$), it is straightforward to show that the detailed balance condition
\begin{equation}
    p(S_t)\,p(S_{t+1}|S_t) = p(S_{t+1})\,p(S_{t}|S_{t+1})
\end{equation}
is satisfied by the equilibrium probability distribution $p(A) = p(B) = p(C) = 1/3$.
The Markov system can be equivalently represented using the variables $X$ and $Y$ alone as in Fig.~\ref{fig:markov_SI}.

Starting from direction $X\rightarrow Y$, we notice that the Eq.~(1) in the main text can be equivalently rewritten as a difference of entropies,
\begin{equation}\label{eq:TE_XtoY_SI}
  \text{TE}_{X\rightarrow Y}(\tau=1) = H(Y_{t+1} \mid Y_t) - H(Y_{t+1} \mid X_t,Y_t)\,,
\end{equation}
where
\begin{subequations}
  \begin{align}
    H(Y_{t+1} \mid Y_t)      & = - \sum_{Y_t,Y_{t+1}} p(Y_{t+1}\mid Y_t)\,p(Y_t)\log p(Y_{t+1}\mid Y_t) \label{eq:entropy_Y_given_Y}\,,                   \\
    H(Y_{t+1} \mid X_t, Y_t) & = - \sum_{X_t,Y_t,Y_{t+1}} p(Y_{t+1}\mid X_t,Y_t)\,p(X_t,Y_t) \log p(Y_{t+1}\mid X_t,Y_t)\,. \label{eq:entropy_Y_given_XY}
  \end{align}
\end{subequations}
By substituting the probabilities appearing in Fig.~\ref{fig:markov_SI} in these equations one finds:
\begin{subequations}
  \begin{align}
    H(Y_{t+1} \mid Y_t)      & = -\frac{2}{3}\bigg(0.3\log(0.3) + 0.7\log(0.7) \bigg) -\frac{1}{3}\bigg(  0.4\log(0.4) + 0.6\log(0.6)\bigg)\,, \\
    H(Y_{t+1} \mid X_t, Y_t) & = -\frac{1}{3}\bigg( 2\times 0.5\log(0.5) + 0.1\log(0.1) + 0.9\log(0.9)                                         \\
                             & \hspace{1.4cm} + 0.4\log(0.4) + 0.6\log(0.6)\bigg)\,. \nonumber
  \end{align}
\end{subequations}
Since $H(Y_{t+1} \mid Y_t) > H(Y_{t+1} \mid X_t, Y_t)$, the Transfer Entropy in direction $X\rightarrow Y$ is positive ($\text{TE}_{X\rightarrow Y}(\tau=1) \simeq 0.07$).

Similarly, the Transfer Entropy in direction $Y\rightarrow X$ can be written as:
\begin{equation}\label{eq:TE_YtoX_SI}
  \text{TE}_{Y\rightarrow X}(\tau=1) = H(X_{t+1} \mid X_t) - H(X_{t+1} \mid X_t,Y_t)\,,
\end{equation}
where
\begin{subequations}
  \begin{align}
    H(X_{t+1} \mid X_t)      & = - \sum_{X_t,X_{t+1}} p(X_{t+1}\mid X_t)\,p(X_t)\log p(X_{t+1}\mid X_t) \label{eq:entropy_X_given_X} \\
    H(X_{t+1} \mid X_t, Y_t) & = - \sum_{X_t,Y_t,X_{t+1}} p(X_{t+1}\mid X_t,Y_t)\,p(X_t,Y_t) \log p(X_{t+1}\mid X_t,Y_t)\,. \label{eq:entropy_X_given_XY}
  \end{align}
\end{subequations}
In this case, all non-zero probabilities appearing in Eq.~(\ref{eq:entropy_X_given_XY}) actually depend on $X_{t+1}$ and $X_t$ only, as the state of the system at time $t$ is fully determined once $X_t$ is known.
This implies $H(X_{t+1} \mid X_t) = H(X_{t+1} \mid X_t, Y_t)$ and, as a consequence, $\text{TE}_{Y\rightarrow X}(\tau=1) = 0$.
We note that if all transition probabilities were set to 0.5, deleting the self-loops of nodes B and C in the original Markov system, an incidental independence on $X_t$ would appear in the non-zero probabilities of Eq.~(\ref{eq:entropy_Y_given_XY}) also in direction $X\rightarrow Y$, resulting in TE$_{X\rightarrow Y}(\tau=1) = 0$.


\section{Langevin system}\label{sec:langevin_SI}

The free energy landscapes depicted in Figs.~3\textbf{d} and 3\textbf{g} in the main text were sampled with an overdamped Langevin dynamics,
\begin{subequations}\label{eq:langevin_dynamics}
    \begin{align}
        \dot{X}(t) &= -\frac{1}{\gamma_X} V'(X,Y) + \sqrt{\frac{2k_B T}{\gamma_X}}\,\eta_X(t)\,, \\
        \dot{Y}(t) &= -\frac{1}{\gamma_Y} V'(X,Y) + \sqrt{\frac{2k_B T}{\gamma_Y}}\,\eta_Y(t)\,,
    \end{align}
\end{subequations}
where $V(X,Y)$ is the free energy of the system, $\gamma_X$ ($\gamma_Y$) is the friction coefficient for $X$ ($Y$), $T$ is the temperature, $k_B$ denotes the Boltzmann constant and $\eta_X(t)$, $\eta_Y(t)$ are independent noise terms satisfying $\langle \eta_X(t)\,\eta_X(t') \rangle = 2 \gamma_X\, k_B\, T \delta(t-t')$ (similarly for $\eta_Y(t)$).
In both cases, we set $k_B\,T = 1$. In Fig.~3\textbf{d}, the friction coefficients were both set to 1, while in Fig.~3\textbf{g} we set $\gamma_X = 1$ and $\gamma_Y = 0.1$, enforcing a slower diffusion in the $X$ direction.

We integrated Eqs.~\ref{eq:langevin_dynamics} with a simple Euler scheme for $N=2500$ independent initial conditions $(X(0),Y(0))$, with $X(0)\sim \mathcal{N}(0,1)$ and $Y(0)\sim \mathcal{N}(0,1)$. 
A time step $dt = 0.01$ was employed and the state was saved every 10 steps.
From each trajectory, the initial $1000$ time points were excluded from the analysis.


\section{Unobserved states in the Markov system}

\begin{figure}[H]
  \centering
  \includegraphics[width=1.\linewidth]{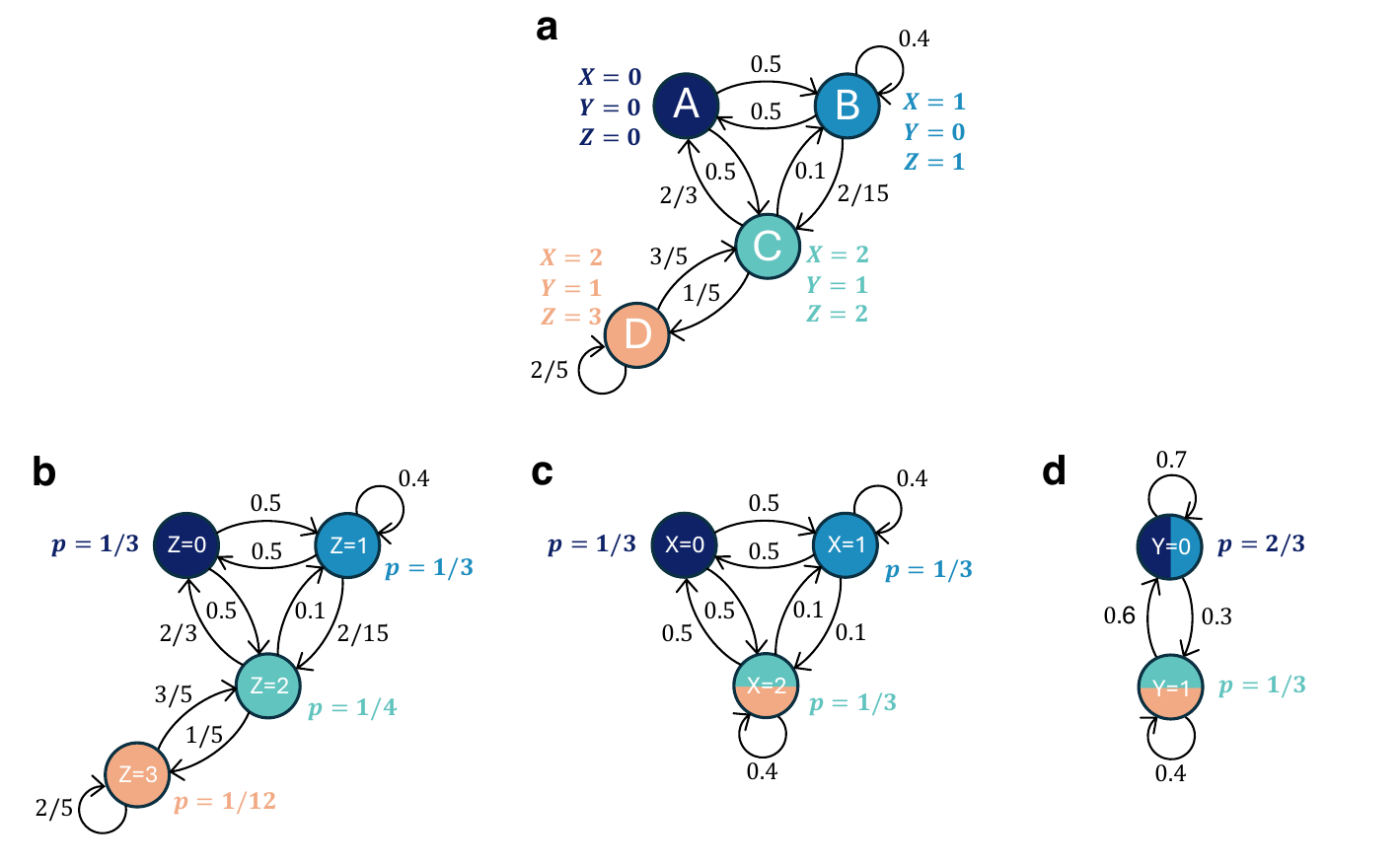}
  \caption{\textbf{a}: Markov system where each state is defined by the triplet $(X, Y, Z)$. Transition probabilities are reported next to the corresponding arrows.
  \textbf{b}: Marginal Markov diagram for $Z$.
  \textbf{c}: Marginal Markov diagram for $X$.
  \textbf{d}: Marginal Markov diagram for $Y$.
  In panels \textbf{b}, \textbf{c} and \textbf{d}, the equilibrium probabilities are reported next to the corresponding states.}
  \label{fig:markov_Z_SI}
\end{figure}

In this Section, we consider a Markov system similar to that introduced in Fig.~3\textbf{a} in the main text, but with an additional state D and a third variable $Z$ which is more informative than $X$ (Fig.~\ref{fig:markov_Z_SI}, panel \textbf{a}).
Here, $Z$ can distinguish all the states of the system (panel \textbf{b}), $X$ has a degeneracy in states C and D (panel \textbf{c}), and $Y$ is degenerate both in C, D and in A, B (panel \textbf{d}).
In the following we show that illustrate this by showing that the bivariate Transfer Entropy satisfies TE$_{X\rightarrow Y}(\tau=1) > 0$, while the multivariate transfer entropy (see Supp. Sec.~\ref{sec:transfer_entropy_SI}) including $Z$ in the conditioning set leads to TE$_{X\rightarrow Y \mid Z}(\tau=1) = 0$.

The equilibrium probability distribution satisfying detailed balance for this Markov-state system is $p(A)=p(B)=1/3$, $p(C)=1/4$ and $p(D)=1/12$.
We note that this system is observationally equivalent to that of Fig.~3\textbf{a} when $Z$ is unobserved, since the two systems share the same transition probabilities between the states distinguished by $X$ and $Y$.

Since the system is observationally equivalent to the three-state system of Fig.~3\textbf{a}, the bivariate Transfer Entropy in direction $X\rightarrow Y$ is the same computed in Supp. Sec.~\ref{sec:markov_system_SI} ($\text{TE}_{X\rightarrow Y}(\tau=1) \simeq 0.07$).

As in the bivariate case, the multivariate Transfer Entropy from $X$ to $Y$ given $Z$ can be decomposed as the difference of two Shannon entropies,
\begin{equation}
    \text{TE}_{X\rightarrow Y \mid Z} = H(Y_{t+1} \mid Y_t, Z_t) - H(Y_{t+1} \mid X_t, Y_t, Z_t)\,,
\end{equation}
where
\begin{subequations}
\begin{align}
    H(Y_{t+1} \mid Y_t, Z_t)  = - \sum_{Y_t,Z_t,Y_{t+1}} p(Y_{t+1}&\mid Y_t,Z_t)\,p(Y_t\mid Z_t)\,p(Z_t)\log p(Y_{t+1}\mid Y_t)\,, \label{eq:entropy_Y_given_YZ}\\
    H(Y_{t+1} \mid X_t, Y_t, Z_t) = - \sum_{X_t,Y_t,Z_t,Y_{t+1}} &p(Y_{t+1}\mid X_t,Y_t,Z_t)\,p(Y_t\mid X_t,Z_t) \label{eq:entropy_Y_given_XYZ}\\
    &\times\,p(X_t,Z_t)\log p(Y_{t+1}\mid X_t,Y_t,Z_t)\,. \nonumber
\end{align}
\end{subequations}

\noindent In all the non-zero probabilities appearing in Eq.~(\ref{eq:entropy_Y_given_XYZ}), the dependence on $X_t$ can be removed, as $X_t$ is a deterministic function of $Z_t$. 
As a consequence, $H(Y_{t+1} \mid Y_t, Z_t) = H(Y_{t+1} \mid X_t, Y_t, Z_t)$ and $\text{TE}_{X\rightarrow Y \mid Z}(\tau=1) = 0$.


\section{Supplementary figures}

\begin{figure}[H]
  \centering
  \includegraphics[width=0.95\linewidth]{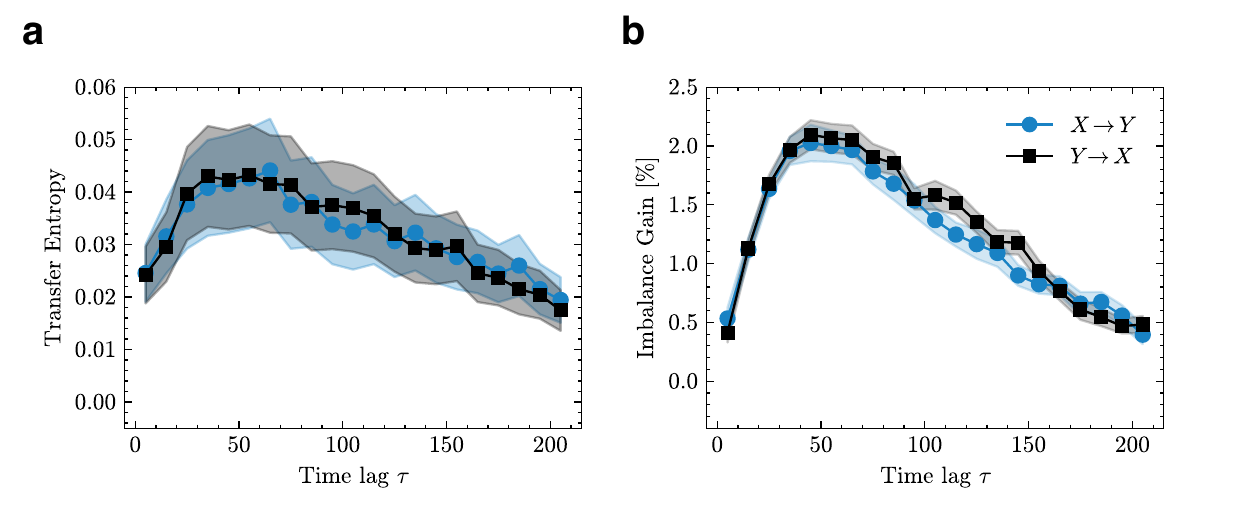}
  \caption{Transfer Entropy (panel \textbf{a}) and Imbalance Gain (panel \textbf{b}) between two variables $X$ and $Y$ described by the same free energy of Fig.~3\textbf{g}, following a Langevin dynamics with identical friction coefficients ($\gamma_X = \gamma_Y = 1$).}
  \label{fig:symm_langevin_SI}
\end{figure}

\begin{figure}[H]
  \centering
  \includegraphics[width=1.0\linewidth]{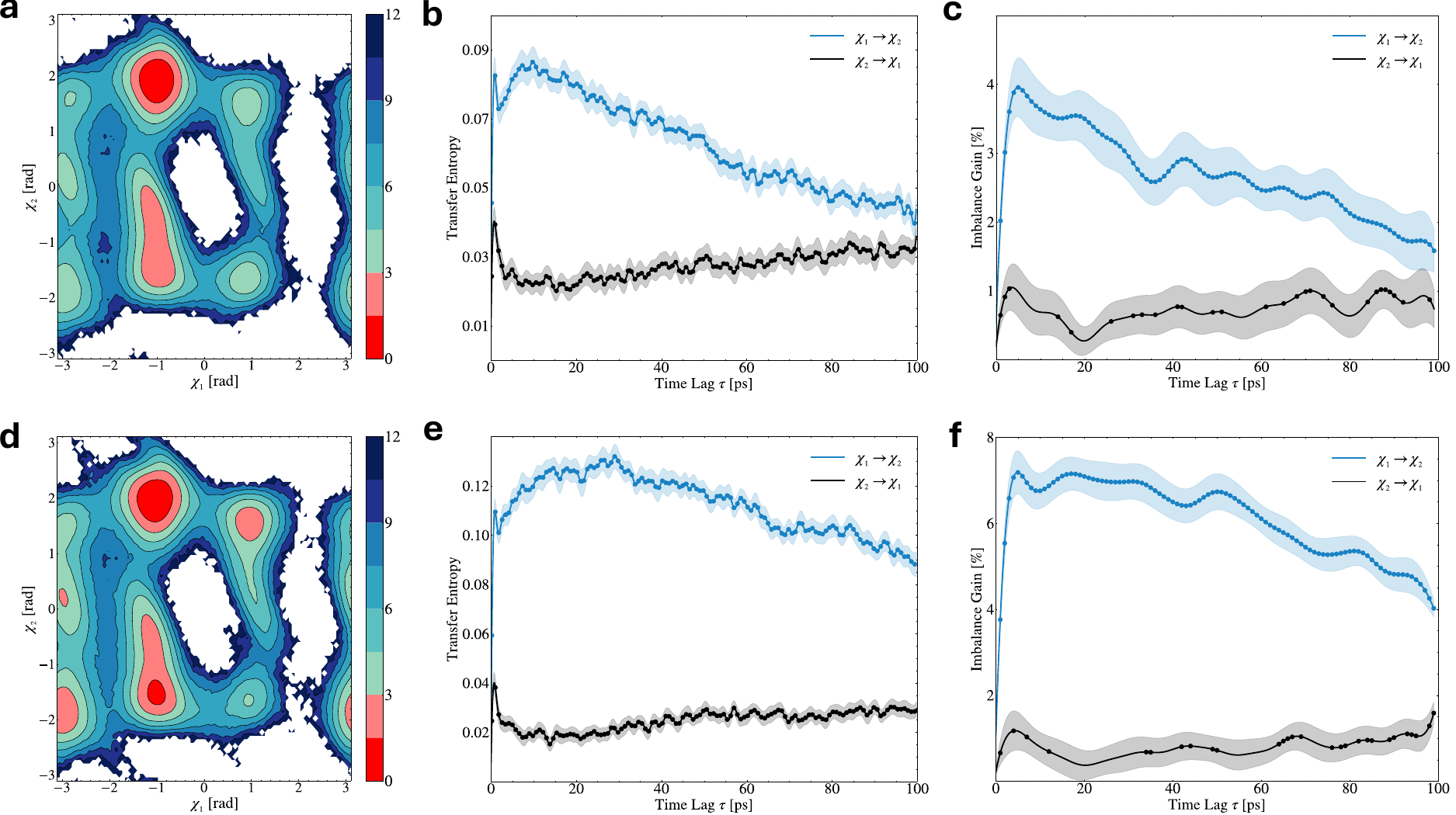}
  \caption{\textbf{a} and \textbf{d}: FES between $\chi_{_1}$ and $\chi_{_2}$ in the GS and ES, respectively. \textbf{b} and \textbf{c}: TE and IG curves for the CVs $\chi_{_1}$ and $\chi_{_2}$ in the GS. \textbf{e} and \textbf{f}: TE and IG curves for the same CVs in the ES. Shaded regions denote error bars over 14 independent estimates, and bold points denote values that are significantly different from 0 according to a t-test ($p < 0.001$).}
  \label{fig:tor1_IG_FES_SI}
\end{figure}

\begin{figure}[H]
  \centering
  \includegraphics[width=1.0\linewidth]{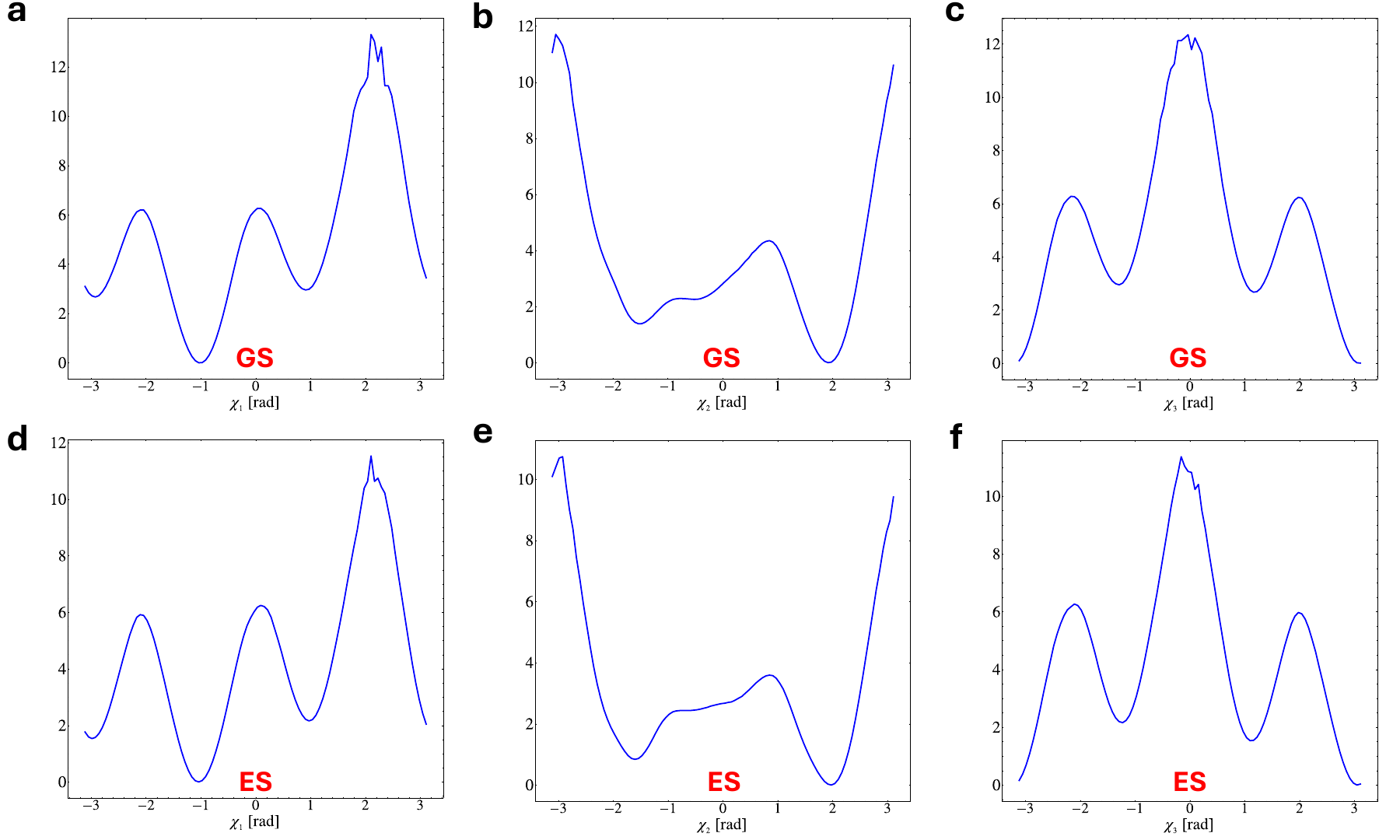}
  \caption{\textbf{a},\textbf{b},\textbf{c}: one-dimensional free energies of the three dihedrals under investigation in this paper - $\chi_{_1}$, $\chi_{_2}$, $\chi_{_3}$, in the GS. \textbf{d},\textbf{e},\textbf{f}: one-dimensional free energies of $\chi_{_1}$, $\chi_{_2}$, $\chi_{_3}$, in the ES.}
  \label{fig:1d_FES_tor_SI}
\end{figure}

\begin{figure}[H]
  \centering
  \includegraphics[width=1.0\linewidth]{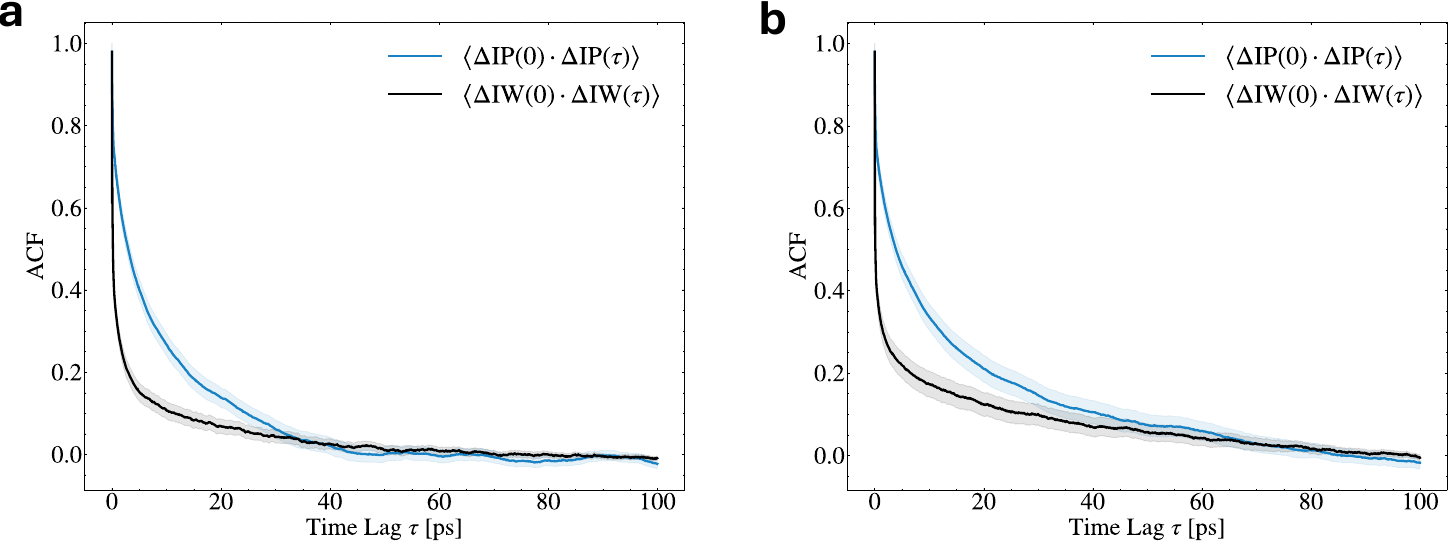}
  \caption{\textbf{a} and \textbf{b}: The autocorrelation functions (ACF) of $\Delta \operatorname{IP}$ (blue) and $\Delta \operatorname{IW}$ (black) in the GS and ES, respectively. Each curve was fit to a single exponential function of the form $\exp(-\frac{t}{\tau})$, where $\tau$ represents the decorrelation time of the corresponding variable. The decorrelation times for $\Delta \operatorname{IP}$ and $\Delta \operatorname{IW}$ in the GS (\textbf{a}) are 7.1 ps and 1.0 ps, respectively. In the ES (\textbf{b}), these values change to 10.7 ps for $\Delta \operatorname{IP}$ and 1.6 ps for $\Delta \operatorname{IW}$.}
  \label{fig:ACF_energies_SI}
\end{figure}

\begin{figure}[H]
  \centering
  \includegraphics[width=1.0\linewidth]{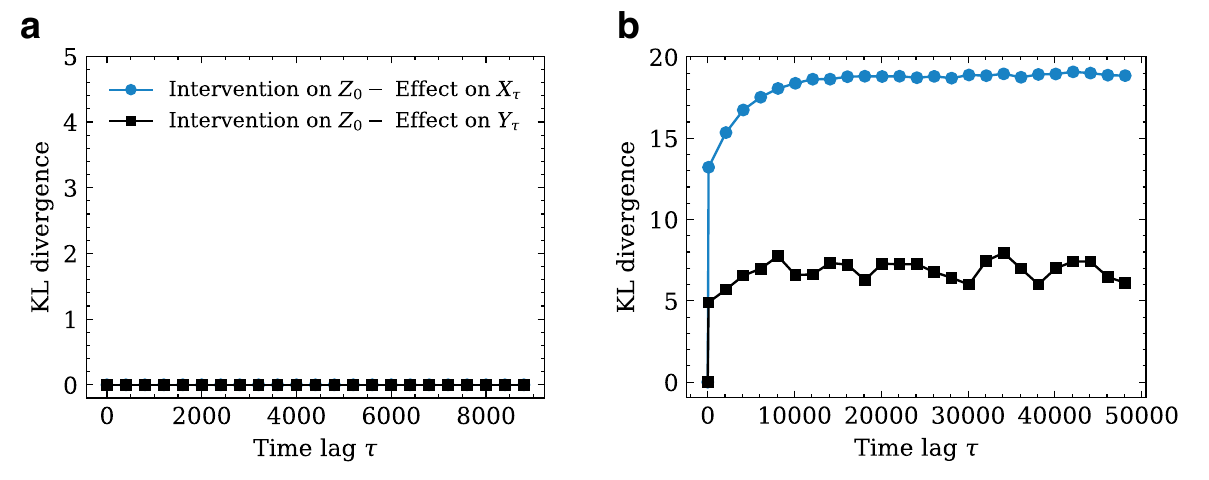}
  \caption{\textbf{a}: Effect of interventions on $Z_0$ for the three-dimensional Langevin systems shown in Fig.~5\textbf{a} of the main text.
  \textbf{b}: Effect of the same interventions for the three-dimensional Langevin system of Fig.~5\textbf{c}.
  The blue and black curves show the KL divergences measuring the effect of the interventions on $X_\tau$ and $Y_\tau$, respectively.
  Specifically, the two curves display $D_{KL}\left[p(X_\tau \mid Z_0 = 0)\, \|\, p(X_\tau \mid Z_0 = 16)\right]$ and $D_{KL}\left[p(Y_\tau \mid Z_0 = 0) \,\|\, p(Y_\tau \mid Z_0 = 16)\right]$.
  The interventional values for $Z_0$ were set to the Z-coordinates of the furthest free energy minima in Fig.~5\textbf{c}.
  }
\end{figure}

\begin{figure}[H]
  \centering
  \includegraphics[width=1.0\linewidth]{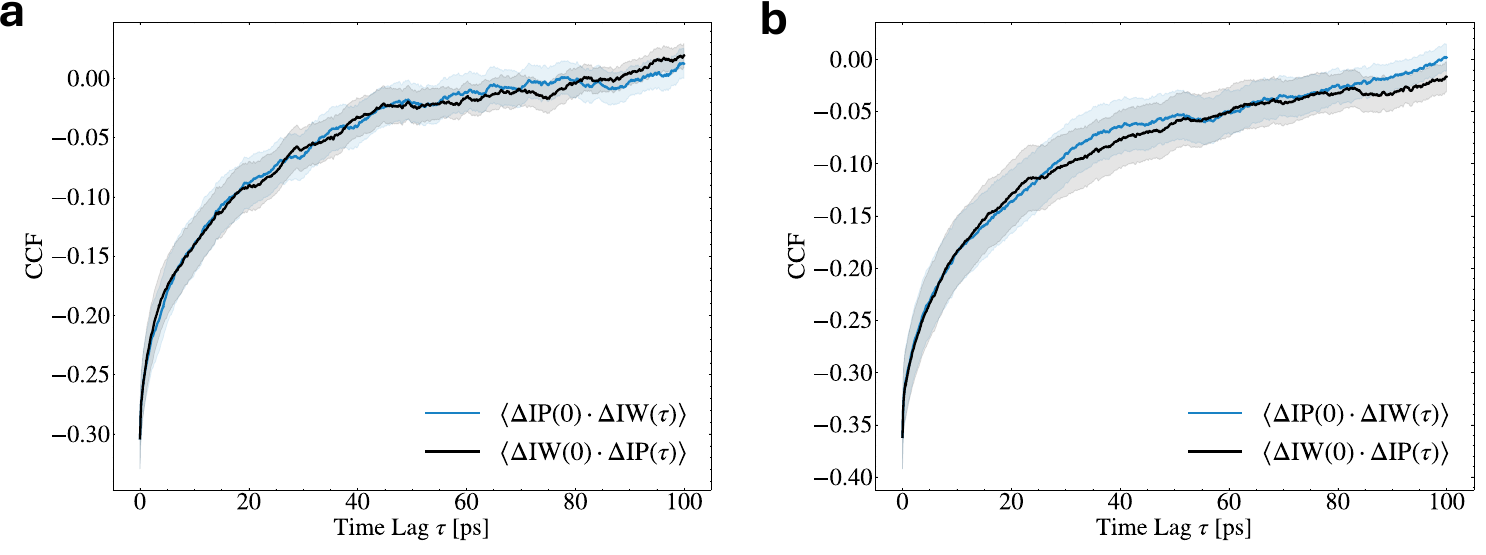}
  \caption{\textbf{a} and \textbf{b}: The cross-correlation functions between $\Delta \operatorname{IP}$ and $\Delta \operatorname{IW}$ in the GS and ES, respectively.  The case where the time lagged variable is $\Delta \operatorname{IW}$ is shown in blue, and where the time lagged variable is $\Delta \operatorname{IP}$ is shown in black.}
  \label{fig:CCF_energies_SI}
\end{figure}

\begin{figure}[H]
  \centering
  \includegraphics[width=1.0\linewidth]{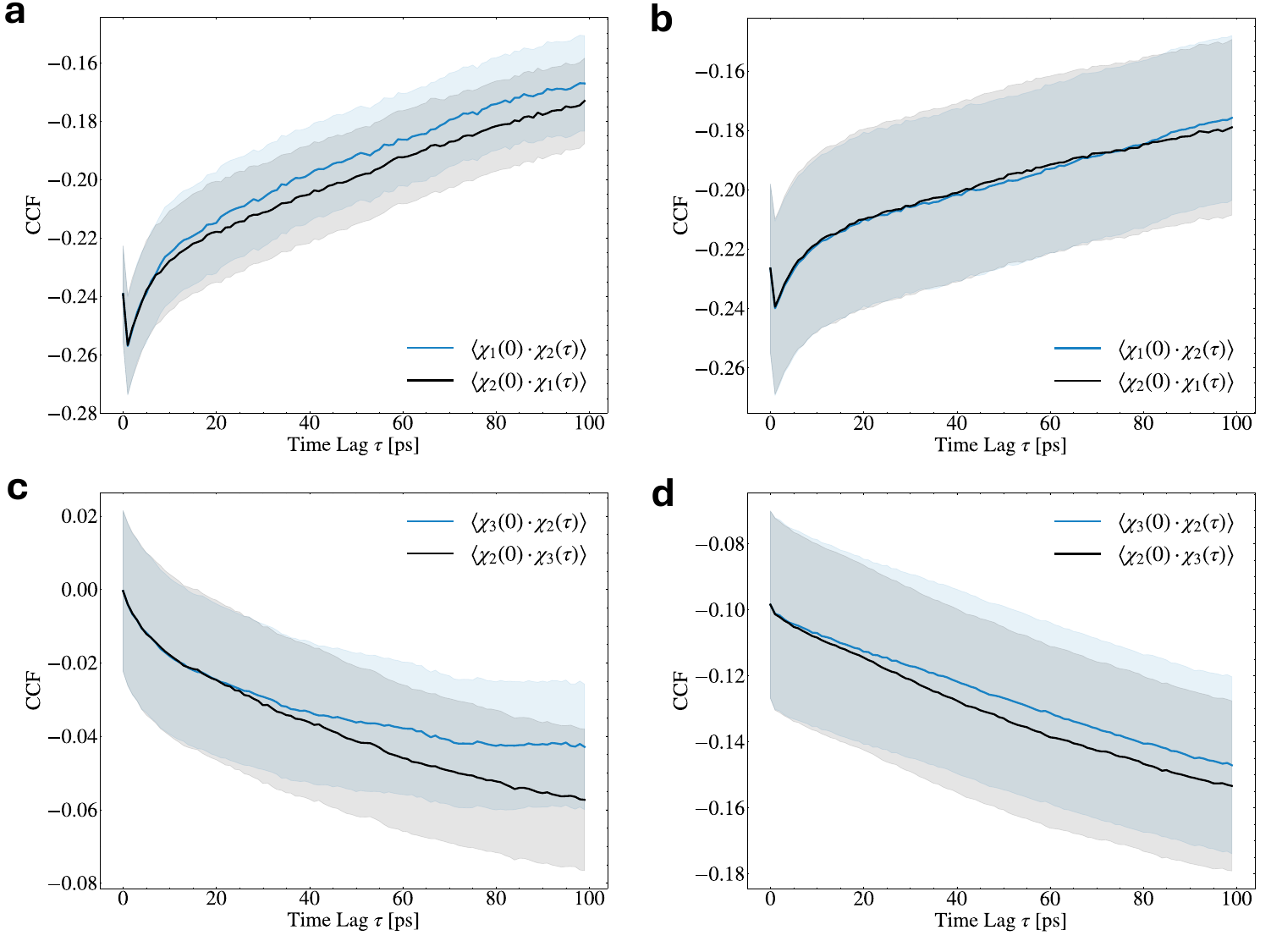}
  \caption{\textbf{a} and \textbf{b}: Cross-Correlation Functions (CCF) between $\chi_{_1}$ and $\chi_{_2}$ in the GS and ES, respectively. \textbf{c} and \textbf{d}: CCFs between $\chi_{_3}$ and $\chi_{_2}$ in the GS and ES, respectively.
  To account for the periodicity of such variables, CCFs of the form $\langle \chi_i(0)\cdot \chi_j(\tau)  \rangle $ were computed as $\frac{1}{2}\langle |\chi_i(0)|_{2\pi}^2 + |\chi_j(\tau)|_{2\pi}^2 - |\chi_i(0) - \chi_j(\tau)|_{2\pi}^2  \rangle $, where $|\chi |_{2\pi} = \min_k |\chi - 2\pi k|$.}
  \label{fig:CCF_tor_SI}
\end{figure}


\bibliography{refs}